\pdfoutput=1 % if your are submitting a pdflatex (i.e. if you have
\documentclass[11pt,a4paper]{article}

\usepackage{jheppub} % for details on the use of the package, please
\usepackage[T1]{fontenc} % if needed
 
\usepackage{amsmath,amsthm,amssymb}
\usepackage{multicol}
\usepackage{soul}
\usepackage{blkarray}
\usepackage{array}

\numberwithin{equation}{section}

\usepackage[utf8]{inputenc}
\usepackage[english]{babel}
\usepackage{textcomp}
\usepackage{amsmath}
\usepackage{mathtools}
\usepackage{gensymb}
\usepackage{multirow}
\usepackage{multicol}
\usepackage{amsbsy}
\usepackage{amssymb}
\usepackage{capt-of}%%To get the caption
\usepackage[T1]{fontenc} % Use 8-bit encoding that has 256 glyphs
\usepackage{physics}
\usepackage{multicol} % Used for the two-column layout of the document
\usepackage{multirow}
\usepackage[table]{xcolor}
\usepackage{lscape}
\usepackage[format=plain,small,labelfont={bf},up]{caption} 
\usepackage[title]{appendix}
\usepackage{graphicx}  % For including images
\usepackage{subcaption}
\def\be{\begin{equation}}
\def\ee{\end{equation}}
\def\bea{\begin{eqnarray}}
\def\eea{\end{eqnarray}}

\def\ie{{\it i.e.~}}

\usepackage{booktabs} % Horizontal rules in tables
\usepackage{float} % Required for tables and figures in the multi-column environment - they need to 
\usepackage{hyperref} % For hyperlinks in the PDF
\usepackage{graphicx} % Para imagens
\usepackage{epstopdf} %para imagens .eps
\usepackage{paralist} % Used for the compactitem environment which makes bullet points with less 
\usepackage{subcaption}
\usepackage[framemethod=TikZ]{mdframed}

\RequirePackage{color}
\usepackage{colortbl}
\definecolor{denim}{rgb}{0.08, 0.38, 0.74}
\hypersetup{
    colorlinks=true,
    linkbordercolor={white},
    linkcolor={denim},
    citecolor={denim},
    filecolor={denim},      
    urlcolor={denim},
}

\begin{document}

\begin{titlepage}
	\vspace*{0.6cm}
	\begin{center}
		{\Huge \color{denim} {\bf 
        Three-Field String Inflation with Perturbative Corrections: Dynamics and Implications}}
		\\[12mm]
        {\LARGE
		\bf Vasileios Basiouris}${^{\diamond}}$~\footnote{E-mail: \texttt{v.basiouris@uoi.gr, v.basiouris@gmail.com}},
		{\LARGE \bf Dibya Chakraborty}${^{\star}}$~\footnote{E-mail: \texttt{dibyac@physics.iitm.ac.in}}
	\end{center}
	\vspace*{0.30cm}
	\centerline{\it
		\Large $\diamond$ \; Physics Department, University of Ioannina}
	\centerline{\it \Large 45110, Ioannina, 	Greece}
    \vspace{0.2cm}
    \centerline{\it
		\Large $\star$ \;Centre for Strings, Gravitation and Cosmology}
	\centerline{\it \Large Department of Physics, Indian Institute of
Technology Madras,} 
\centerline{\it
		\Large Chennai 600036, India}
	\vspace*{1.20cm}
   %Three-field Inflation utilising Perturbative Corrections in String Theory 
	\begin{abstract}

    In this work, we construct an explicit string motivated example of three-field inflation in a related, yet distinct from the recently discovered perturbative large volume scenario (pLVS). In this set up, large volume is ensured by the interplay between the effects of $\alpha^{\prime 3}$, logarithmic loop and higher derivative $F^4$ corrections. After addressing full moduli stabilization, we move on to a detailed analysis of a three-field model of inflation while choosing a suitable canonical basis for the fields. Our model differs from previous setups in three key aspects: first, the interaction between subleading corrections that drive full moduli stabilization is different compared to pLVS for instance; second, the volume of the underlying Calabi-Yau is of "weak-Swiss-cheese" type; and third, in our three-field inflationary scenario, the second Hubble slow-roll parameter consistently dominates over the first by several orders of magnitude --- primarily due to the steepness of the potential. When this hierarchy is aided by a brief deviation from slow-roll, an enhancement in the primordial power spectrum can be naturally achieved. We have obtained two stages of inflation: the first stage mostly occurs when one of the inflatons roll along the steepest direction ---completing $55$ efolds of inflation. Once the rolling inflaton falls off a ridge of the multi-field potential, the other two fields become active, giving us a true multi-field behavior in the second stage -- adding few more efolds of inflation. Inflation ends when all three of them roll towards the global minimum and settles there. We confirm this behavior by introducing the non-planar torsion in the inflationary trajectory -- this quantity becomes non-trivial in the second stage of inflation. Finally, we calculate the cosmological observables, which align with the Planck data, and discuss potential directions for future research. 
			
	%In this work, we have addressed moduli stabilization in the framework of a recently found -- well motivated Perturbative Large Volume Scenario, combining the effects of $\alpha^{\prime 3}$, logarithmic loop and higher derivative $F^4$ corrections. Focusing on the K{\"a}hler moduli sector, a detailed analysis is presented regarding the minima of the effective scalar potential, while the basis of the normalized fields is explicitly computed. Given the above, the inflationary potential is studied ....

	\end{abstract}
	
 \end{titlepage}

%\maketitle
\vspace{0.2cm}
{\color{denim}\hrule}
\vspace{0.2cm}

\tableofcontents
\vspace{0.5cm}
{\color{denim}\hrule}
\vspace{0.5cm}

\section{Introduction}

Cosmic inflation refers to a rapid expansion of spacetime occured in the early Universe \cite{Guth,Linde81,AS82}, which smoothed out any fluctuations in spacetime, resulting in a homogeneous and isotropic cosmos. Our current understanding of the Universe and its evolution is based on the six-parameter \(\Lambda\)CDM model. The model lacks an explanation for the horizon problem and the observed flatness of our Universe. Inflation can provide an explanation for the above two, besides, it can also give rise to the seed for the structure formation we observe today. In addition, the data from Planck \cite{Planck:2018jri} are fully consistent with the simplest inflationary  scenario  being the leading mechanism to account for the observed anisotropies in the Cosmic Microwave Background (CMB) radiation. Specifically, they align with two strong predictions of inflation: a nearly scale-invariant power spectrum of density perturbations and a stochastic background of gravitational waves.\par

According to the simplest inflationary scenario, inflation is driven by a scalar field minimally coupled to gravity, whose potential energy dominates over its kinetic energy, causing it to slowly roll over its potential. Once the kinetic energy begins to dominate, inflation comes to an end, providing sufficient spacetime expansion to resolve the horizon problem. However, recently proposed consistentcy conjectures \cite{Vafa:2005ui,Obied:2018sgi, Agrawal:2018own, Garg:2018reu,Palti:2019pca} imposed on the 4D effective field theory (EFT) derived from a quantum theory of gravity, strongly disfavors vanilla single field inflation. Although, it was hinted upon in \cite{Achucarro:2018vey} that multi-field inflation with non-trivial field space metric might be able to alleviate this tension.\par

Moreover, in this decade and the next, various experiments are being launched with aim of detecting the B-mode polarisation in the CMB induced by primordial gravitational waves. They are: CMB-S4 experiment \cite{CMB-S4:2016ple}, CLASS \cite{2014SPIE.9153E..1IE}, LiteBIRD \cite{Matsumura:2013aja}, and the Simon Observatory \cite{SimonsObservatory:2018koc}. Moreover, upcoming Large Scale Structure surveys, including DESI \cite{DESI:2016fyo}, LSST \cite{LSST}, Euclid \cite{Amendola:2016saw}, and SKA \cite{Camera:2014bwa}, may also detect primordial non-Gaussianities, that may unravel potential interactions during the inflationary era \cite{Meerburg:2019qqi}. Both from theoretical and experimental point of view, it thus becomes essential to go beyond the single field vanilla scenario for inflation. In particular, multifield models are generic in supergravity and string theory realisations.\par

Four-dimensional EFT emerging after dimensional reduction of a superstring theory features various scalar fields, and one/some of them might act as natural candidate/s to drive inflation. This forms the foundation of our three-field inflation model. However, these constructions also have a price to pay; most of the time, the existence of too many scalar fields becomes unavoidable, leading to the so-called cosmological moduli problem (see \cite{Cicoli:2023opf} and the references therein). One also needs to stabilize the scalar fields to prevent the appearance of fifth forces. Our aim in this paper is to display a concrete three-field inflationary scenario where the scalar potential is constructed out of the underlying low-energy effective string theory setup, in particular from a type-IIB flux compactification where the extra dimensions are compactified on a K3 fibred Calabi-Yau (CY). The first step towards string theory model building is to stabilize the scalar fields. Fortunately, string theory provides sufficient tools (such as $Dp$ branes pierced with magnetized fluxes) to help achieve this goal.\par

The scalar fields stemming from the closed string sector of type-IIB flux compactifications are of three types: complex-structure $(U^i)$ related to the shapes of the CY, K\"ahler moduli $(T_{\alpha})$ related to the size of the CY, and an Universal scalar field known as axio-dilaton $(s)$. Vacuum expectation value (vev) of the imaginary part of axio-dilaton determines the string coupling $g_s$. 
After dimensional reduction, in 4D supergravity, the scalar potential can be expressed in terms of a holomorphic superpotential and a real K\"ahler potential. The Type-IIB  superpotential depends on the complex structure moduli and dilaton, however it is independent of the K\"ahler moduli, resulting in the so-called \emph{no-scale structure}. Hence, to stabilize the K\"ahler moduli, one has to go beyond and add further quantum corrections, such as perturbative corrections to the K\"ahler potential and non-perturbative effects at the level of superpotential. Note that the complex structure and axio-dilaton are stabilized at tree level in the presence of fluxes and integrated out. \par

The most notable candidate models --- addressing full moduli stabilization are the KKLT \cite{Kachru:2003aw,Polchinski:2015bea, Gautason:2019jwq,Hamada:2019ack,Linde:2020mdk, Randall:2019ent,Blumenhagen:2019qcg,Andriot:2020wpp, Abe:2005rx,Carta:2019rhx} model and the Large Volume Scenario (LVS) \cite{Conlon:2005ki, Balasubramanian:2005zx,Burgess:2003ic, Cremades:2007ig}. The former utilises non-perturbative effects to be added in a holomorphic superpotential of the Gukov-Vafa-Witten form. The latter addresses the moduli stabilization by a careful interplay between perturbative effects at the level of real K\"ahler potential and non-perturbative effects to the superpotential. These balance in LVS setup helps to achieve an exponentially large volume of the underlying 6D Calabi-Yau manifold spanned by the K\"ahler moduli. \par

KKLT and LVS being the main two examples to address full moduli stabilization, recently another class of string vacua also emerged \cite{Antoniadis:2019rkh,Antoniadis:2018hqy,Basiouris:2020jgp, Basiouris:2021sdf, Leontaris:2022rzj} where it was shown that exponentially large volume can be achieved without the need of non-perturbative effects --- this type of construction is known as pLVS. Non-perturbative effects are very restrictive and very specific towards the underlying CY geometry, additionally they may need to satisfy several constraints such as Witten’s unit arithmetic genus condition \cite{Witten:1996bn}. Therefore, if a construction ensures a large internal volume without relying on these effects, it becomes more well-motivated. A recent study \cite{Bera:2024zsk} have provided a global embedding of the pLVS framework and the robustness of the same against various subleading corrections, including leading-order corrections in inverse string tension, higher-derivative corrections, log-loop corrections from intersecting space-filling \( D7 \) brane stacks, D-term effects, and Kaluza-Klein as well as winding-type loop corrections.

%It is worth mentioning, also, the fact that both scenarios, are focusing on the string theory's background geometry (the type IIB theory) and the various extended objects like D-branes in order to argue for the emergence of the various corrections, where their presence also influence the uplifting mechanism used to result into a dS vacuum. More recent studies on similar issues can be found in the following list \cite{Cicoli:2016xae, Cicoli:2017axo,AbdusSalam:2020ywo,2309.11697,Cicoli:2023njy,Bera:2024ihl} \par
The present work studies the addition of perturbative effects such as leading $\alpha'^3$ correction (BBHL term), corrections of log-loop type and from higher derivative terms, in a weak Swiss-cheese CY. We start deriving the multi-field scalar potential and then disucss the stabilization of the scalars putting special emphasis on the higher derivative $F^4$ effect and the log-loop corrections. Then we move on to discuss their inflationary implications. Inflation takes place in two stages: initially along the steep direction of the potential, and then along the remaining two directions as all the fields move towards their global minimum from an initially displaced position. At this initially displaced position in the scalar potential, the minimum along one of the scalars disappears, producing a steep ridge in that direction. The inflaton then falls in the ridge and oscillates --- in this stage the other two fields move relatively slower, giving rise to a quassi-single field behavior in the first stage. In the later stage, the field that fell into the ridge moves relatively slowly. Since the fields are canonical, the non-planar motion and deviation from a geodesic occur only when the field temporarily falls into a ridge and oscillates there. The oscillation can not only give rise to high turning rate and torsion, but it can also induce sizable entropic perturbation through transient deviation from slow-roll. The backreaction of the entropic modes to the adiabatic modes may produce a significant enhancement in the primordial power spectrum signalling a copious production of primordial blackholes (PBH). In this lore, it is also useful to study the effect of slow-roll violation in the secondary induced gravitational waves. We aim to come back to the implications of slow-roll violation in a future companion paper.\par %Finally, we discuss the inflationary predictions and conclude with the possible future research directions such as the copious production of primordial blackholes (PBH).%DC needs to change this paragraph.

%\textbf{A paragraph needed for the various inflationary models in this field. In this paragraph a few things about axions and reheating should be stated with the following refs included. In addition, in similar stringy constructions, studies are focusing on the other aspects of cosmological evolution like the dark radiation and its correlation to the dark matter \cite{Leedom:2024qgr,Cicoli:2018cgu,Cicoli:2022uqa,Gu:2018akj,Cicoli:2013oba,Basiouris:2024buq,Allahverdi:2014ppa,Cicoli:2012aq,Higaki:2012ar,Angus:2014bia}.}\par 

The layout of the paper is as follows: in section \ref{section2}, we present a short overview of the quantum corrections to be added in the K{\"a}hler potential along with their string origin and their effect in the stabilization of the K\"ahler moduli. We end the section by displaying several examples of de Sitter vacua. In section \ref{section3}, we introduce the multi-field setup where we introduce several concepts like turning rate and torsion. We also discuss their implications at the level of linear perturbation theory. Section \ref{section 4} mainly covers our detailed three-field inflationary analysis. After performing several consistency checks stemming from supergravity and the hierarchy of mass scales, we move on to checking the inflationary predictions dubbed in the value of spectral index, spectral tilt, and the amplitude of power spectrum. We also discuss how turning and torsion can affect them. We conclude with our future directions in \ref{section5}. 
 %In addition, we discuss the implications of aforementioned corrections in the moduli stabilization, where the next to leading order term is sufficient to create a potential fore the small divisor. In section4., we summarize the important ingredients for the multi-field inflation case. In section 4. we embedded an inflationary paradigm in the current framework, where we discuss explicitly the implications of logarithmic corrections on the observables and the reason why a multi-field scenario is preferred. Finally, the concluding remarks are summarized in section 5. \textbf{Vasilis: Needs modifications regarding the inflation part.}
 
\section{Towards moduli stabilization with perturbative effects}\label{section2}

%\subsection{Origin of the logarithmic loop corrections}
The perturbative string loop corrections has been recently highlighted in \cite{Antoniadis:2018hqy} due to their major role in the moduli stabilization mechanism without the need for a non-perturbative effects like the Large Volume Scenario \cite{Cicoli:2008gp}. In this section, we start by summarizing the world-sheet corrections which are proportional to the Euler characteristic \( \chi \) of the compactification, followed by an analysis of the origin of string loop corrections within the geometric framework of $D7$ branes in type IIB string theory. After presenting a detailed derivation of the scalar potential, we shall showcase several examples of dS vacua where the correct mass hierarchy remains preserved.\par
\subsubsection*{$\alpha'^3$ corrections}

$\alpha'^3$ corrections \cite{Becker:2002nn} are added at the level of the  K{\"a}hler potential to stabilize the volume modulus $\mathcal{V}$. It acts as an additive factor to the tree-level K\"ahler potential, giving us:
\begin{equation}
    \mathcal{K}=-2\log(\mathcal{V}+\dfrac{\xi}{2g_s^{3/2}})~.
\end{equation}
 This constant shift $\xi$ of the volume is given in terms of the Euler characteristic of the manifold as: \begin{equation}
    \xi= - \dfrac{\zeta(3)}{4(2\pi)^3}\chi~.
\end{equation}
%{\color{red}Starting from the $\alpha^{\prime}$ corrections \cite{Becker:2002nn}, it is proven that the 4D dilaton field in Einstein frame can be redefined in terms of the 10D dilaton as:
%\begin{equation}
%    e^{-2\phi_4}=e^{-2\phi_{10}}(\mathcal{V}+\xi)=e^{-\frac{1}{2}\phi_{10}}(\mathcal{V}+\hat{\xi}),
%\end{equation}
%
%The first term in the parentheses stands for the overall internal volume, which can be expressed in terms of the imaginary part of K{\"a}hler moduli $T^i$.
%sAs for the second term, it represents corrections to the K{\"a}hler potential, where readily one can see that a shift $\hat{\xi}$ in the volume will absorb this term. This constant shift is given by \cite{Becker:2002nn}: 
%
\subsubsection*{Log-loop corrections}

Now, the origin of string loop corrections can be traced back to higher derivative terms proportional to $R^4$ that appearing in the effective action due to multi-graviton scattering of the ten dimensional theory. To make the discussion more precise, we can express the effective action as \cite{Antoniadis:2019rkh}:
\begin{eqnarray} 
{\cal S }_{\rm grav}&=& \frac{1}{(2\pi)^7 \alpha'^4} \int\limits_{M_{4} \times {{\cal X}_6}} e^{-2\phi} {\cal R}_{(10)} - \frac{\chi}{(2\pi)^4 \alpha'} \int\limits_{M_{4}} \left(-2\hat{\xi}(3) e^{-2\phi}  \pm 4\hat{\xi}(2) \right) {R}_{(4)}\,,  
\label{IIBAction} 
\end{eqnarray} 
where $\phi$ is the string dilaton and 
 $ { R}_{(4)}$ denotes the Ricci scalar. in four dimensions and the $\pm$ signs refers to the different type of string theory, type IIA/B theory respectively \cite{Antoniadis:1997eg,Antoniadis:2002tr,Antoniadis:2003sw}. In the last step, $\chi$ is defined as :
\begin{equation}	
\chi= 	\frac{3}{4\pi^3}\int\limits_{{\cal X}_6} R\wedge R\wedge R~\cdot 
\label{Euchi}
\end{equation}
From the above, it is observed that an additional localized Einstein-Hilbert (EH) is induced in four dimensions with a coefficient proportional to $\chi$. These terms correspond to localized gravity vertices in the bulk (in particular at points where $\chi \neq 0$), where their effect is to emit Kaluza-Klein states and gravitons. We assume a background with  three intersecting $D7$ branes, where we expect interactions between the closed string modes and KK states in the bulk space. These interactions exhibit in general a logarithmic behavior in the large volume limit transverse to the $D7$-branes. The new contributions to the ${R}_{(4)}$ term which are going to be moduli dependent, can be summarized by the following formula \cite{Basiouris:2020jgp}:
\begin{eqnarray}
\delta=\frac{4\zeta(2)}{(2\pi)^3}\chi \int_{M_4} \left(1-\sum_k e^{2\phi}T_k \ln(R^k_{\bot}/\mathtt{w})\right)\,R_{(4)}~.\label{allcor}
\end{eqnarray}
$T_k$ is the tension of a $D7$-brane, %DC: check this : Vasilis: 1909.10525 3.15 equation or 2007.15423 equation 13.
and $R_{\bot}$ stands for the size of the two-dimensional world volume in each brane. Moreover, $\mathtt{w}$ parametrizes a `width' utilized as an ultraviolet
cutoff for the graviton KK modes.\par
The low-energy dynamics of the four-dimensional effective supergravity theory, emerging from type IIB superstring compactifications on CY orientifolds, can be described by a holomorphic superpotential (\( W \)) and a real K\"ahler potential (\( K \)). The tree-level superpotential and K\"ahler potential are functions of various chiral coordinates. These chiral coordinates are constructued by writing them as a combination of a modulus with a set of RR axions, such as:
\begin{equation}
    U^i=v^i-iu^i,\qquad s=c_0+i e^{-\phi},\qquad
    T_{\alpha}=c_{\alpha}-i\,\tau_{\alpha},
\end{equation}
where $\phi$ is the universal modulus known as dilaton, related to string coupling as $g_s=e^{\langle\phi\rangle}$; $u^i$ is the saxion and $v^i$ is the axion of the complex-structure modulus, and $\tau_{\alpha}$'s are the Einstein frame four-cycle volume moduli defined as, $\tau_{\alpha}=\partial_{t^{\alpha}}\mathcal{V}$, $\mathcal{V}$ is the overall volume of the CY manifold in Einstein frame expressed as $\mathcal{V}=\dfrac{1}{3!}\kappa_{\alpha\beta\gamma}t^{\alpha} t^{\beta} t^{\gamma}$. Finally, $c_0$ and $c_{\alpha}$'s are the universal RR axion and RR four-form axions, respectively.

The addition of these corrections changes the tree-level K{\"a}hler potential to:
\begin{equation}
    K= - \log(-i(s-\bar{s})) -2\log\left[\mathcal{V}+\dfrac{\xi}{2}\left(\dfrac{s-\bar{s}}{2i}\right)^{3/2}+ \left(\dfrac{s-\bar{s}}{2i}\right)^{-1/2} \delta\right],\quad \delta=\hat{\eta} \log[\tau_i],
\end{equation}

\subsection{The geometrical setup}

Taking into account the above discussions, we now turn to the derivation of the scalar potential and discuss the CY geometry. Fibre inflation model, proposed more than a decade ago in \cite{Cicoli:2008gp} is promising with respect to the embedding of inflation in string compactifications, since they provide a natural solution to the $\eta-$problem of inflation \cite{Cicoli:2011zz}. Different variations of fibre inflation models have been proposed, where combinations of string loop corrections, Kaluza-Klein and winding loops are considered, leading into inflationary potentials with a sufficient plateau-like form.\par
In all of these models, the presence of non-perturbative effects to stabilize the blow-up model and the overall volume were common. Recently, another aspect of fibre inflation is also proposed where the interplay of perturbative corrections assured a large volume \cite{Bera:2024ihl} as well as full moduli stabilization. The absence of non-perturbative effects eliminates the need for an exceptional del Pezzo divisor. The novelty of this setup is that the usual drawback of fibre inflation where the field space excursion was bounded by K\"ahler cone conditions were alleviated. The use of perturbative corrections in the context of inflationary models can also be found in \cite{Bansal:2024uzr} where the perturbative effects were used in the stabilization of the blow-up mode. Later, from the perspective of brane-antibrane inflation, these effects are also proven useful in \cite{Cicoli:2024bwq}.  \par

In this paper, we follow a similar approach as \cite{Bera:2024ihl} but consider a different form of the volume, resembling that of K3-fibred Calabi-Yau models like fibre inflation. However, as will become clear in the following subsections, our inflationary predictions deviate significantly from those of the fibre models because we do not utlize non-perturbative effects to stabilize the overall volume and the blow-up modulus. To proceed with our analysis, the compactified space volume is given by:
\begin{equation}
\mathcal{V}=f_{3/2}(\tau_i)-\sum_j^{N_{rg}}\lambda_j\tau_j^{3/2}~ i=1,..,N,
\end{equation}
 where the $f_{3/2}$ is a homogeneous function of the K\"ahler moduli. Despite lacking a complete global embedding for our approach, we are going to adapt the following volume form: 
%\noindent Given the above analysis, the resulting compactification space is described by three K{\"a}hler moduli, where the non-linearity of the space is still preserved.
%
\begin{equation}
\mathcal{V}=\dfrac{1}{\sqrt{2\alpha}}\sqrt{\tau_7}\tau_6 -\dfrac{1}{3}\tau_1^{3/2}~,\label{volume2}
\end{equation}
where $\alpha$ is a model dependent parameter, parametrizing the D-term constraint. Given this form, we cannot a priori characterize the hierarchy between the moduli since the log-loop corrections might affect the anticipated result for $\tau_1\ll \tau_{6,7}$. Nevertheless, we are obliged to respect the requirement of having a large volume. It would be worth searching through the Kreuzer-Skarke list of Calabi-Yau manifolds embedded in toric varieties \cite{Kreuzer:2000xy,Altman:2014bfa} for some representative examples where our geometric configuration of intersection $D7$ could lead to large log-loop corrections. $\tau_1$ is the blow-up mode, $\tau_7$ is the usual fibre modulus --- the volume of the K3 manifold, and $\tau_6$ is the volume of the $\mathbb{P}^2$ base. Their cumulative effect ensures that the model remains within the large volume scenario during inflation by adjusting their values.

\subsection{The effective scalar potential and moduli stabilization}
%\textbf{We aim to stabilise the blow-up mode witht he help of log-loop, $F^4$ and alpha cubed corrections -- write this}

Having this volume-form as a starting point, our analysis will be focused on the addition of BBHL term and logarithmic corrections in the K{\"a}hler potential without considering any non-perturbative corrections. The large volume in this model is ensured by a cumulative effect of the perturbative corrections. The K{\"a}hler potential upon the insertion of the these corrections can be written as:
\begin{equation}
\mathcal{K}=-\log\left[\dfrac{s-\bar{s}}{i}\right]-2\log\left[\mathcal{V}+\dfrac{\xi}{2}\left(\dfrac{s-\bar{s}}{2i}\right)^{3/2}+\left(\dfrac{s-\bar{s}}{2i}\right)^{-1/2}\sum_{i}\hat{\eta}_i\log(\tau_i)\right]\label{Kahler}
\end{equation}
where $\tau_i$ stand for the four-cycle moduli, related to the volume of the CY as $\mathcal{V}\sim \tau^{3/2}$, $\xi$ denotes the $\alpha^{\prime}$ correction-term and the logarithmic corrections $\sim \hat{\eta} \log(..)$ are induced along each four-cycle volume moduli. One important last observation we have to make is the fact that the stabilization will completely depend on perturbative corrections. The non-perturbative corrections, which will impose stringent limitation in our construction, could be prevented by considering certain world-volume fluxes that lift fermionic zero modes leading to zero contribution from these corrections \cite{Green:1997di,Bianchi:2011qh,Cvetic:2012ts,Blumenhagen:2007sm}. Taking this into account, in this work, we will explore the case where $
\tau_1$ no longer remains at small values, but it is stabilized to large values but never taking us away from the domain of large volume.% With respect to the large volume expansion, our effective potential has to ensure that the values of the compactified volume has to be positive and the expansion over $\xi/\mathcal{V}$ stays under control. \par
 Despite the fact that this approach diverges from the typical paradigm of fibred models, our proposal focuses on the implications on the logarithmic correction on the weak-Swiss-cheese CY geometry of \eqref{volume2}. \par
Regarding the complex structure moduli $U^i$ as well as the axion-dilaton, they are stabilized at tree-level by the supersymmetric conditions:
\begin{equation}
D_{U^i}W=0=D_{\bar{U}^i}\bar{W}, \;D_{s}W=0=D_{\bar{s}}\bar{W}~.
\end{equation}
We will also set:
\be
\mathcal{W}_0=\left\langle\int_{X_6}G_3\wedge \Omega\right\rangle
\ee
where $G_3=F_3+i\,sH_3$ with $F$ and $H$ denote RR and NSNS 3-form fluxes respectively. $\Omega$ is the holomorphic $(3,0)-$form which depends on the complex sturcture moduli $U^i$. Since the scalar potential is parametrized by the three moduli ($\tau_1,\tau_7,\tau_6$), we would like to trade one of them in order to introduce the volume $\mathcal{V}$ in the computations. This reparametrization helps us to perform the large volume expansion. So, in order to include the overall volume in our computations, we solve the equation \eqref{volume2} with respect to one modulus e.g. $\tau_6$,
\begin{equation}
\tau _6= \frac{\sqrt{2\alpha } \left(\tau _1^{3/2}+3 \mathcal{V}\right)}{3 \sqrt{\tau _7}}~.
\end{equation}\label{t6}
Using the above definition, we are going to present the necessary computations needed to define the effective F-term scalar potential. Moreover, following \cite{Leontaris:2022rzj}, the functions in the K{\"a}hler potential can be written as follows:
\begin{equation}
 Y=\mathcal{V}+\dfrac{\hat{\xi}}{2} + \hat{\eta}  \log \left(\frac{2 \alpha \tau_7}{9}  \left(\tau_1^{5/2}+3 \mathcal{V} \tau_1\right)^2\right)~.\label{Ydef}
\end{equation}
where we have used the fact that the axio-dilaton $s$ and the free parameters of the model $(\hat{\xi}, \hat{\eta})$ are recasted as:
\begin{align}
s=c_0+i e^{-\phi},\; d=e^{-\phi}\Rightarrow \hat{\xi} =\xi\; d^{3/2},\; \hat{\eta}= \hat{\eta}\; d^{-1/2},\;\;d=\dfrac{1}{g_s}~.
\end{align}
Imposing these redefinitions in \eqref{Ydef}, the K{\"a}hler potential can be rewritten as:
\begin{equation}
 \mathcal{K}=-\log\left[\dfrac{s-\bar{s}}{i}\right]-2\log\left[Y\right]~.
\end{equation}
The scalar potential can be computed by the $\mathcal{N}=1$ supergravity formula:
\begin{equation}
e^{-\mathcal{K}}V_k=K^{A\bar{B}}(D_AW)(D_{\bar{B}}\bar{W})-3|W|^2,\;\;W=\mathcal{W}_0~.\label{ScPot}
\end{equation}
 The first term of the scalar potential \eqref{ScPot} will be displayed below, where we expand in terms of $\mathcal{O}(\hat{\eta}, \hat{\xi}/\mathcal{V})$ keeping only the first order terms:
\begin{equation}
K^{A\bar{B}}(D_AW)(D_{\bar{B}}\bar{W})=3g_s\mathcal{W}_0^2\bigg( 1+\dfrac{\hat{\xi}-16\hat{\eta}+\hat{\eta} w}{4\mathcal{V}}+\dfrac{\hat{\xi} \hat{\eta} (1-2w)}{2\mathcal{V}^2}\bigg) + \mathcal{O}\left(\dfrac{\hat{\xi}}{\mathcal{V}^n}\right),
\end{equation}
 where we define $w= \log \bigg(\dfrac{2\alpha \tau_7}{9}(3\mathcal{V}\tau_1+\tau_1^{5/2})^2\bigg)$. It is obvious that the second term in \eqref{ScPot} cancels the first term above, and by turning off $\hat{\eta} \rightarrow 0$, it results to:
\begin{equation}
e^{-\mathcal{K}}V_{eff}=\dfrac{3 g_s\mathcal{W}_0^2 \hat{\xi}}{4\mathcal{V}}+ \mathcal{O}\left(\dfrac{\hat{\xi}}{\mathcal{V}^n}\right)~.\label{BBHL1}
\end{equation} 
Based on the above, the effective scalar potential can be easily derived and it is shown below:
\begin{equation}
V_{eff}=\dfrac{3g_s \mathcal{W}_0^2(\hat{\xi}-16\hat{\eta}+\hat{\eta} w)}{4\mathcal{V}^3}+ \mathcal{O}\left(\dfrac{\hat{\xi}}{\mathcal{V}^n},\hat{\eta}^n\right)~.\label{EffPott}
\end{equation}
As discussed in \cite{Cicoli:2017axo,Cicoli:2008gp}, it is a unique feature of these models to preserve one flat direction. In \cite{Cicoli:2008gp,Cicoli:2017axo}, their approach is to stabilize first the internal volume and the blow-up divisor with an interplay of perturbative and non-perturbative effects, while the $\tau_7$ remains flat. Our approach differs from the aforementioned due to the fact that the logarithmic corrections will be made sufficient to perform the stabilization while keeping track of the remaining flat direction. The previous studies of this kind of approach requires only keeping terms that scales as $\sim \frac{1}{\mathcal{V}^3}$ , but now we shall prove that even terms proportional to $\sim \frac{\hat{\eta} \hat{\hat{\xi}}}{\mathcal{V}^4}$ could be significant. Firstly by inspecting the effective potential \eqref{EffPott}, we stabilize with respect to to $\mathcal{V}$ while considering $\mathcal{V}\gg \tau_1$:
\begin{equation}
\partial_{\mathcal{V}} V_{eff}=\frac{9 \mathcal{W}_0^2 g_s}{4 \mathcal{V}^4} \left(-\hat{\eta}  \log \left(\frac{2 \alpha  \tau
   _7}{9}\right)-\hat{\xi} +2 \hat{\eta}  \left(\frac{\mathcal{V}}{\tau _1^{3/2}+3
   \mathcal{V}}+8\right)-2 \hat{\eta}  \log \left(\tau _1^{5/2}+3 \tau _1
   \mathcal{V}\right)\right)=0,\notag 
\end{equation}
\begin{equation}
\Rightarrow\mathcal{V}\cong \frac{e^{\frac{1}{6} \left(-3 \log \left(\frac{2 \alpha \tau_7}{9}\right)-\frac{3 \hat{\xi} }{\hat{\eta} }+50\right)}-\tau _1^{5/2}}{3 \tau _1}~.
\end{equation}
By substituting the above minimum in the scalar potential, the result is:
\begin{equation}
V_{eff}|_{\mathcal{V}_{min}}=-\frac{27 \sqrt{2} \hat{\eta}\;  \tau _1^3\;  \mathcal{W}_0^2 \left(\alpha  \tau
   _7\right){}^{3/2} e^{\frac{3 \hat{\xi} }{2 \hat{\eta} }} g_s}{\left(\sqrt{2}
   \sqrt{\alpha  \tau _1^5 \tau _7} e^{\frac{\hat{\xi} }{2 \hat{\eta} }}-3
   e^{25/3}\right){}^3},
\end{equation}
where we can see that $\tau_{1,7}$ directions still remain flat. To remedy this, we now include the next to leading order term: 
\begin{equation}
V_{eff}=\dfrac{3g_s\mathcal{W}_0^2(\hat{\xi}-16\hat{\eta}+\hat{\eta} w)}{4\mathcal{V}^3}-\dfrac{3g_s\mathcal{W}_0^2 \hat{\eta} \hat{\xi}(w)}{\mathcal{V}^4}+ \mathcal{O}\left(\frac{\hat{\xi}^n}{\mathcal{V}^n},\hat{\eta}^n\right)~.
\end{equation}
This new correction will not modify the minimum w.r.t. the volume (at least to a degree that would lead to destabilization), since it is suppressed by $\sim \frac{1}{\mathcal{V}^4}$. Nevertheless, it would slightly lift the contributions to the transverse directions unraveling the importance of quantum corrections to the stabilization of the flat direction. So, the effective potential at $\mathcal{V}_{min}$ is:
\begin{equation}
V_{eff}|_{\mathcal{V}_{min}}=\frac{27 \tau _1^4 \mathcal{W}_0^2 g_s \left(\frac{3 \sqrt{2} \hat{\eta} 
   e^{\frac{25}{3}-\frac{\hat{\xi} }{2 \hat{\eta} }}}{\tau _1 \sqrt{\alpha  \tau _7}}+12
   \hat{\xi}  (3 \hat{\xi} -50 \hat{\eta} )-2 \hat{\eta}  \tau _1^{3/2}\right)}{4
   \left(e^{\frac{1}{6} \left(-3 \log \left(\frac{2 \alpha  \tau
   _7}{9}\right)-\frac{3 \hat{\xi} }{\hat{\eta} }+50\right)}-\tau _1^{5/2}\right){}^4},
\end{equation}
where minimizing with respect to $\tau_7$, there exist one extremum at:
\begin{equation}\label{eq:2.24}
\tau_7=\frac{9 e^{\frac{50}{3}-\frac{\hat{\xi} }{\hat{\eta} }}}{2 \alpha  \left(12 \hat{\xi} \tau
   _1+\tau _1^{5/2}\right){}^2}~.
\end{equation}
Given the above extremum, we achieve a saddle at:% while there is a flat direction in terms of a combination of the $\tau_1,\tau_7$ moduli.
\begin{equation}
V_{eff}|_{\mathcal{V}_{min}}^{\langle\tau_7\rangle}=\frac{3 \mathcal{W}_0^2 (\hat{\xi} -16 \hat{\eta} ) g_s}{256 \hat{\xi} ^3}>0~. 
\end{equation}
Despite the fact that the effective potential is positive at the minimum given $\hat\xi>16\hat\eta$, not all the directions are stabilized. As we can see in \eqref{eq:2.24}, the extremum along $\tau_7$ depends on $\tau_1$ which is yet to be fixed. Thus, we are obliged to investigate the possibility of additional corrections, whose effects could achieve this. The following correction can also be induced as an addition to the leading order $\alpha'^3$ corrections. They are summarized in the following contributions: %\textbf{shift KK and winding to the other tex file.}

\noindent\textbf{Higher-derivative string corrections:} They appear at the level of $\alpha'^3$ corrections multiplying an $F^4$ term and they modify the 4D EFT scalar potential as \cite{Ciupke:2015msa}:
\begin{equation}
    V_{F^4}=-k^2 \dfrac{|\lambda| \mathcal{W}_0^4}{g_s^{3/2}\mathcal{V}^4}\Pi_i t^i,
\end{equation}
where $\Pi_i$ is a topological number depending on the intersection of the divisors and $\lambda$ is unknown combinatorial factor. 

\subsubsection*{Inflationary potential and moduli stabilization}
In the present model, we are going to utilize this subleading correction and study their implications both on the stabilization procedure and on the inflationary analysis: 
\begin{align}\label{V_exact}
    V_{eff}=\dfrac{3g_s\mathcal{W}_0^2(\hat{\xi}-16\hat{\eta}+\hat{\eta} w)}{4\mathcal{V}^3}-\dfrac{3g_s\mathcal{W}_0^2 \hat{\eta} \hat{\xi}(w)}{\mathcal{V}^4} + \dfrac{c}{\mathcal{V}^4} (t_1 +t_6+t_7) + V_{up}+ \mathcal{O}\left(\dfrac{1}{\mathcal{V}^5}\right), 
\end{align}
where the $c$ parameter quantifies the overall factor of $F^4$ corrections depending on the string coupling $g_s$ and the fluxes $\mathcal{W}_0$. In addition, we have added a small constant $V_{up}$ which will be used to obtain an almost Minkowski/dS vacuum. Some efforts towards carefully computing this uplifting term have been analyzed in \cite{Cicoli:2016xae,Cicoli:2024bxw}. Now, for concreteness, we should stress the fact that winding loop corrections and KK corrections should also be considered in a concrete model as have been stressed in \cite{Bera:2024ihl,Bera:2024zsk}, but since we are agnostic about the value of their coefficients, in the current work we only study the case where only $F^4$ corrections is enough to stabilize the flat directions and produce a dS-shallow potential. \par
The relation between the two-cycle moduli and the four-cycle moduli is summarized below and it can be used to express the $F^4$ correction:
\begin{align}
    t_1=\sqrt{\tau_1}, \quad t_7=\dfrac{1}{\sqrt{2\alpha}}\dfrac{\tau_6}{\sqrt{\tau_7}},\quad t_6=\dfrac{1}{\sqrt{2\alpha}}\sqrt{\tau_7} ~.
\end{align}
\begin{align}
    V_{F^4}= \dfrac{c}{\mathcal{V}^4}\bigg(\frac{\sqrt{\tau _7}}{\sqrt{2} \sqrt{\alpha }}+\sqrt{\tau _1}+\frac{\tau _1^{3/2}+3 \mathcal{V}}{3 \tau _7}\bigg), \quad c=-k^2 \dfrac{|\lambda| \mathcal{W}_0^4}{g_s^{3/2}}\Pi_i~.
\end{align}
Based on this addition, one can proof that the flat direction along $\{\tau_1,\tau_7\}$ are lifted. In order to connect this section with the inflationary analysis in the next section, it is easier to express the potential in the normalized field basis spaneed by $(\varphi_1,\varphi_2,\varphi_3)$. To do so, we need to write down the leading order terms of the K{\"a}hler metric, which are:
\begin{align}
K_{ij}=\left(
\begin{array}{ccc}
 \frac{1}{4 \tau _7^2} & 0 & 0 \\
 0 & \frac{1}{2 \tau _6^2} & 0 \\
 0 & 0 & \frac{\sqrt{\alpha  \tau _7}}{4 \sqrt{2} \sqrt{\tau _1} \tau _6 \tau _7} \\
\end{array}
\right)~.
\end{align}
Given the above form, the kinetic term are now written as:
\begin{equation}
    K_{ij}\partial T_i \partial \bar{T_j}=\sum_i \dfrac{1}{2}(\partial \varphi_i)^2+...~.\end{equation}
%\sum_i
% \dfrac{\partial T_i \partial \bar{T_i}}{(T_i+\bar{T}_i)^2}=
The new relation between the old and new bases where the off-diagonal entries are found small in the $\varphi\sim\mathcal{O}(1)$ limit and ignored, are the following:
\begin{align}
    &\tau _7= e^{\sqrt{2} \varphi_1}, \quad \tau _6= e^{\varphi_2}, \\
    &\tau _1= \left(\frac{3^{4} \varphi_3^{4} e^{ \left(\sqrt{2} \varphi_1+2 \varphi_2\right)}}{2^{5}\alpha }\right)^{1/3},\quad \mathcal{V}= \frac{\left(4-3\varphi_3^2\right) e^{\frac{\varphi_1} {\sqrt{2}}+\varphi_2}}{4 \sqrt{2} \sqrt{\alpha }}~.\label{newdefs}
\end{align}
Finally, we display the scalar potential in the new basis as:
\begin{small}
\begin{align}
    V_{eff}=& \,V_{up} -\frac{1024 \alpha ^2 \hat{\xi}  \hat{\eta}\;  g_s \mathcal{W}_0^2 e^{-2 \sqrt{2} \varphi_1-4 \varphi_2} \left(\log \left(\frac{9^4 \varphi_3^8}{2^{10} \alpha
   ^2}\right)+8 \sqrt{2} \varphi_1+10 \varphi_2\right)}{\left(4-3 \varphi_3^2\right)^4}\notag \\
   &-\frac{32 \sqrt{2} \alpha ^{3/2} g_s \mathcal{W}_0^2\; e^{-\frac{3 \varphi_1}{\sqrt{2}}-3 \varphi_2} \left(\hat{\eta}  \left(\log \left(\frac{9^4 \varphi_3^8}{2^{10} \alpha ^2}\right)+8 \sqrt{2} \varphi_1+10 \varphi_2-48\right)+3 \hat{\xi} \right)}{\left(3 \varphi_3^2-4\right)^3}\notag \\
   &+\frac{512\;\; 2^{1/6}\; \alpha ^{3/2}\; c\;\; e^{-\frac{5 \varphi_1}{\sqrt{2}}-4 \varphi_2} \left((9 \alpha\; \varphi_3^2\; 
   e^{2 \sqrt{2}\varphi_1+\varphi_2})^{1/3}+2^{1/3} \left(e^{\sqrt{2} \varphi_1}+e^{\varphi_2}\right)\right)}{\left(4-3 \varphi_3^2\right)^4}   ~.\label{exact}
\end{align}
\end{small}
Note that, the potential carries an asymptote along the $\varphi_3$ direction at $\varphi_3=\sqrt{4/3}$. The implications of this will have a clear effect on the inflationary analysis as will be shown in section \ref{section 4}. \par
Due to the complexity of the above potential, we can reduce its form by considering an expansion with respect to a small parameter $y$. One can readily see in the above form, that there is a factor of $4-3\varphi_3^2$, which can be recasted to:

\begin{align}
    \varphi_3^2=\dfrac{y+4}{3}~.\label{redef1}
\end{align}
 Given the definitions in \eqref{newdefs}, we can see that $\phi_3$ is bounded, since we want a positive volume. One can deduce from that the $y$ parameter has to be small in order to satisfy the above constrain. So, we can perform a an expansion with respect to this parameter. The leading order terms of this expansion results into:
\begin{align}
    V_{eff}^{appr}=&\,\frac{32\; \alpha ^{3/2}\;  e^{-\frac{5 \varphi_1}{\sqrt{2}}-4 \varphi_2}}{3 y^4}  \bigg[48 \sqrt{2} c \left(e^{\sqrt{2} \varphi_1}+e^{\varphi_2}\right)+4\times 2^{5/6}\; (3\alpha)^{1/3}\; c (y+12)\; e^{\frac{1}{3} \left(2 \sqrt{2} \varphi_1+\varphi_2\right)}  \notag \\
   &-3 \mathcal{W}_0^2\; y g_s e^{\sqrt{2} \varphi_1+\varphi_2} \left(2 \hat{\eta}  \left(\sqrt{2} (-\log (2 \alpha )+5 \varphi_2-24)+8 \varphi_1\right)+3
   \sqrt{2} \hat{\xi} \right)\notag \\
   &-96 \sqrt{\alpha }\; \hat{\xi}  \hat{\eta}  e^{\frac{\varphi_1}{\sqrt{2}}} \mathcal{W}_0^2 g_s \left(-2 \log (2 \alpha )+8 \sqrt{2} \varphi_1+10 \varphi_2+y\right)\bigg]+ V_{up}~.
\end{align}
We could further simplify our formula considering the regime where the $\alpha$ parameter is small $\alpha \ll 1$. The simpler formula for the effective potential is then given by:
\begin{align}
    V_{eff}^{appr}=&\,\frac{32 \alpha ^{3/2} e^{-\frac{5 \varphi_1}{\sqrt{2}}-4\varphi_2}}{3 y^4}  \bigg[ 4 \sqrt{2} c \left(12 \left(e^{\sqrt{2} \varphi_1}+e^{\varphi_2}\right)+  (y+12) (6\alpha e^{ \left(2 \sqrt{2} \varphi_1+\varphi_2\right)})^{1/3}\right)- \notag \\
   & -\mathcal{W}_0^2 y\; g_s\; e^{\sqrt{2} \varphi_1+\varphi_2} \left(-\sqrt{2} \hat{\eta}  \log \left(64 \alpha ^6\right)+9 \sqrt{2} \hat{\xi} +6 \hat{\eta}  \left(8
   \varphi_1+\sqrt{2} (5 \varphi_2-24)\right)\right)  \bigg]+ V_{up}~.\label{eff}
\end{align}
From this expression, one can easily minimize the potential with respect to the variable $y$. This can be approximated as:
\begin{align}\label{expression_y}
    y\cong \frac{64 \sqrt{2} c\;\; e^{-\sqrt{2} \varphi_1-\varphi_2} \left( (6\alpha e^{ \left(2 \sqrt{2} \varphi_1+\varphi_2\right)})^{1/3} +e^{\sqrt{2} \varphi_1}+e^{\varphi_2}\right)}{\mathcal{W}_0^2 g_s \left(\sqrt{2} \hat{\eta}  (-\log \left(64 \alpha ^6)+30 \varphi_2-144\right)+9 \sqrt{2} \hat{\xi} +48 \hat{\eta}  \varphi_1\right)}~.
\end{align}
By recalling the redefinition in \eqref{redef1}, the minimum along $\varphi_3$ direction is determined as:
\begin{align}\label{expression_varphi3min}
    \varphi_{3,min}^2=\dfrac{1}{3}(y+4)~.
\end{align}
As for the other two perpendicular directions, we could substitute the minimal value of $y$ in the effective scalar potential of equation \eqref{eff}, which leads to the following minima along $(\varphi_1,\varphi_2)$:
\begin{align}
  &  \varphi_{2,min}\cong \frac{1}{30} \left(\log \left(64 \alpha ^6\right)-\frac{9 \hat{\xi} }{\hat{\eta} }-24 \sqrt{2}\varphi_1+30\ W_{0/-1}\left(\frac{ 2^{9/5}
   e^{\frac{1}{30} \left(\frac{9 \hat{\xi} }{\hat{\eta} }+54 \sqrt{2} \varphi_1-184\right)}}{3 \alpha^{1/5} }\right)+184\right),\\
  &  \varphi_{1,min}\cong \frac{6 \hat{\eta}  \log (\alpha )-9 \hat{\xi} +2 \hat{\eta}  \left(144+15 \log \left(\frac{13}{5}\right)+\log (8)\right)}{54 \sqrt{2} \,\hat{\eta} }~.
\end{align}
To prove the validity of our approximate formula of the potential in \eqref{eff} and the exact potential in \eqref{exact}, we are going to sketch the potentials in the vicinity of the analytical and numerical global minimum. 
\begin{figure}[H]
    \centering
    \includegraphics[scale=0.5]{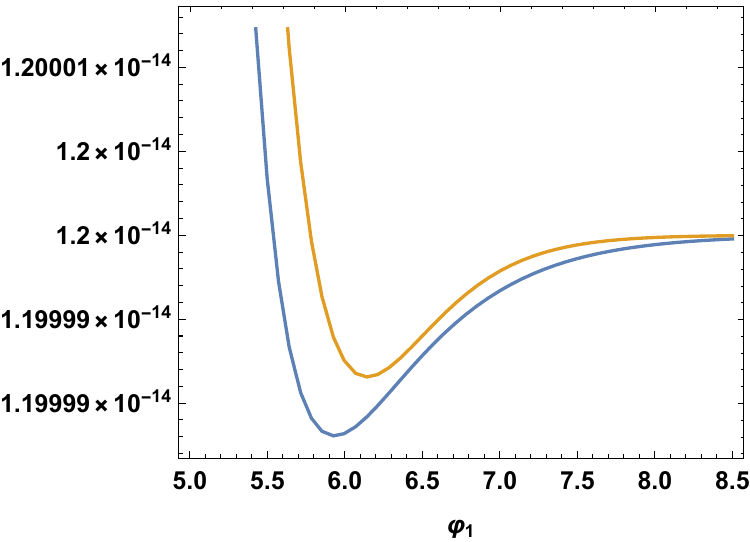} 
    \qquad\includegraphics[scale=0.5]{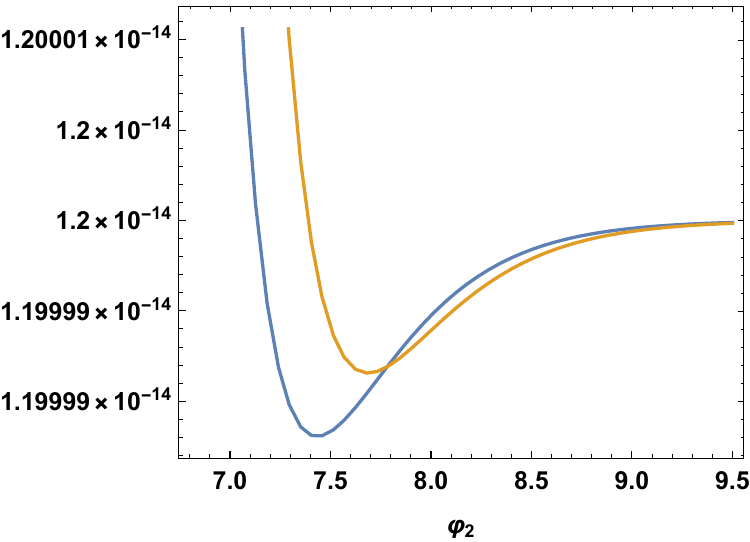}
    \begin{center}\includegraphics[scale=0.65]{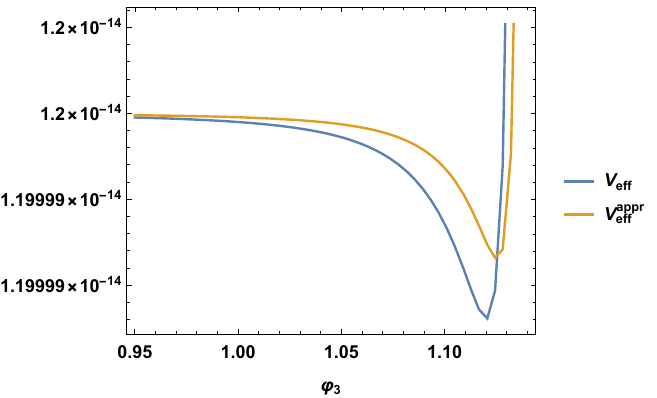}
    \end{center}\caption{Comparison between analytical and numerical computation of minimum along $\varphi_1,\varphi_2$ and $\varphi_3$ directions. The parameters used in these plots correspond to Model 1 of Table \ref{tab2}.}
\end{figure}
%\textit{\textbf{Vasilis: Dibya, are these plots created with the parameters you used for the inflationary analysis ?. Maybe we have to make the font bigger in the plots.}}
%In the above plots, we plot the exact potential \eqref{exact} against our approximate potential \eqref{eff}.
%It is readily seen that our approximation can be justified, since it can effectively describe the dynamics of the moduli in this region of the parameter space. We also confirm the existence of this global minimum through dynamical evolution of the fields presented in section \ref{section 4}.
It is readily seen that our approximated and numerical solutions match modulo tiny departure. Note that, for the inflationary analysis studied in section \ref{section 4}, we work with the potential with the exact form as demonstrated in \eqref{V_exact}.

\subsubsection*{Mass-scales}

Let us now move on to discussing several mass scales associated with the underlying type-IIB flux compactification. The first mass scale of interest is the string scale which for consistency of the low-energy string theory should be much less than the Planck scale with the relation\footnote{Note that, here instead of using natural units, we have reintroduced the Planck scale to understand the hierarchy between various mass scales.} $M_s=\frac{g_s^{1/4}\sqrt{\pi}}{\sqrt{\mathcal{V}}}M_{pl}$. Next, we define our Kaluza-Klein scale associated with the bulk \ie $\tau_{\text{bulk}}^{3/2}=\mathcal{V}$, $M_{KK}=\frac{\sqrt{\pi}}{\mathcal{V}^{2/3}}M_{pl}$. Gravitino mass is given by $m_{3/2}=\sqrt{\kappa}\frac{|W_0|}{\mathcal{V}}M_{pl}$. $m_{3/2}$ sets the scale of all complex structure moduli, the dilaton and the K\"ahler moduli. Another important consistency requirement is $m_{3/2} < M_{KK}$ , so that the gravitino mass
of the 4d EFT is not integrated out. In order to trust the underlying EFT, we need to maintain the following hierarchy of mass scales,
\begin{equation}\label{hierarchy1}
    m_{3/2}<M_{KK}<M_s<M_{pl}.
\end{equation}

\subsubsection*{Example dS-vacua}

In the following tables, several numerical examples are presented, where the values of the free parameters are displayed along with the values of the moduli/fields at their corresponding minimum. 
%\textbf{{\color{blue}I think I convinced myself that phi3 is the heaviest among them}} \textit{\textbf{Vasilis: The correct hierarchy is the following $\phi_3<\phi_2<\phi_1$. Please check the update notebook.}}
%\textbf{{\color{blue}Then how do we defend $\varphi_3$ rolling before anyone else. Can you convince me too? if you are right then my hypothesis is valid only in thevlimit if $\epsilon_V^{\varphi_3}$ is bigger than others. I confirm the same by showing these plots to you.}} \textit{\textbf{Vasilis: The plots confirm your claim or mine ?}}
%\begin{figure}[ht]
 %   \centering
    % Subfigure 1
%    \begin{subfigure}[b]{0.45\textwidth}
 %       \centering
 %       \includegraphics[width=\textwidth]{plots/eps_V_fields.pdf}
       % \caption{(a))}
%    \end{subfigure}
%    \hfill
    % Subfigure 2
 %   \begin{subfigure}[b]{0.45\textwidth}
  %      \centering
   %     \includegraphics[width=\textwidth]{plots/eta_V_fields.pdf}
       % \caption{}
 %   \end{subfigure}
    %\caption{Overall caption for the figure.}
 %   \label{fig:main}
%\end{figure}

\begin{table}[H]
\begin{center}
\begin{tabular}{|c|c|c|c|c|}
\hline
  & \cellcolor[gray]{0.9} $\tau_1$ & \cellcolor[gray]{0.9}$\tau_7$ & \cellcolor[gray]{0.9}$\tau_6$ & \cellcolor[gray]{0.9}$\mathcal{V}$   \\
\hline 
 \cellcolor[gray]{0.9} \text{Model 1} &$1.35\times 10^5$ & $4403$ & $1690$ & $1\times 10^6$\\
 \hline
  \cellcolor[gray]{0.9} \text{Model 2} & $3.56\times 10^5$ &  $7120$ &  $2732$ &  $2\times 10^6$\\
 \hline
  \cellcolor[gray]{0.9} \text{Model 3} &$3.79\times 10^6$ & $44636$ & $17191$ & $1\times 10^8$\\
 \hline
 \end{tabular}
 \begin{tabular}{|c|c|c|}
 \hline
\cellcolor[gray]{0.9} $\varphi_1$& \cellcolor[gray]{0.9}$\varphi_2$ & \cellcolor[gray]{0.9}$\varphi_3$ \\
 \hline
  $5.93$ & $7.43$ & $1.12$ \\
 \hline
  $6.27$ &  $7.91$ & $1.14$ \\
 \hline
 $7.57$ & $9.75$ & $1.13$ \\
 \hline
\end{tabular}
\quad \begin{tabular}{|c|c|c|c|c|c|c|c|}
 \hline
 &\cellcolor[gray]{0.9} $m_{\langle\varphi_1\rangle}$ & \cellcolor[gray]{0.9}$m_{\langle\varphi_2\rangle}$ & \cellcolor[gray]{0.9}$m_{\langle\varphi_3\rangle}$ & \cellcolor[gray]{0.9}$m_{3/2}$ & \cellcolor[gray]{0.9}$M_{\text{KK}}$ & \cellcolor[gray]{0.9}$M_s$  \\
 \hline
 \cellcolor[gray]{0.9}  \text{Model 1} &$3.42\times 10^{-8}$ & $4.63\times 10^{-10}$ & $2.17\times 10^{-10}$ & $5.72 \times 10^{-8}$ & $1.72 \times 10^{-4}$ & $1.73 \times 10^{-4}$ \\
 \hline
  \cellcolor[gray]{0.9}  \text{Model 2} & $3.2\times 10^{-8}$ &  $2.01\times 10^{-10}$ &  $9.27\times 10^{-11}$ &  $3.46 \times 10^{-8}$ & $1.1 \times 10^{-4}$ & $1.19 \times 10^{-4}$ \\
 \hline
 \cellcolor[gray]{0.9}  \text{Model 3} &$2.32\times 10^{-11}$ & $2.28\times 10^{-13}$ & $1.05\times 10^{-13}$ & $2.87 \times 10^{-10}$ & $7.72 \times 10^{-6}$ & $9.51 \times 10^{-6}$ \\
 \hline
\end{tabular}
\end{center}
\caption{Three different numerical solutions and the corresponding mass scales.}
\label{tab1}
\end{table}%

\begin{table}[H]
\begin{center}
\begin{tabular}{|c|c|c|c|c|c|c|c|c|}
\hline
  &\cellcolor[gray]{0.9}$g_s$ & \cellcolor[gray]{0.9}$|\mathcal{W}_0|$ & \cellcolor[gray]{0.9}$\hat{\xi}$  & \cellcolor[gray]{0.9}$|\hat{\eta}|$ & \cellcolor[gray]{0.9}$\alpha$ & \cellcolor[gray]{0.9}$|\lambda|$  & \cellcolor[gray]{0.9}$V_{up}$\\
\hline
 \cellcolor[gray]{0.9} \text{Model 1} & $ 10^{-4}$ & $30$ & $11.5$ & $0.5$ & $2\times 10^{-5}$ & $0.09$ & $1.2\times 10^{-14}$\\
 \hline
  \cellcolor[gray]{0.9} \text{Model 2} & $8.5\times 10^{-5}$ & $38$ & $13.5$ & $0.5$ & $5\times 10^{-5}$ & $0.04$ & $2.23\times 10^{-20}$\\
 \hline
 \cellcolor[gray]{0.9} \text{Model 3} & $ 10^{-5}$ & $50$ & $19.5$ & $0.5$ & $1\times 10^{-6}$ & $0.078$ & $2.86\times 10^{-26}$\\
 \hline 
\end{tabular}
\end{center}
\caption{The free parameters used for the examples presented in table \ref{tab1}.}% In the aforementioned models, the sign of $\lambda$ is negative in order to achieve a stable vacuum.} 
\label{tab2}
\end{table}%

A very recent study \cite{Lacombe:2024jac} showed that the higher derivative corrections are still under consideration regarding the correct embedding in the supersymmetric field theories and their dimensional reduction. In this work it is argued that the inclusion of all the higher derivative terms derived from the ten dimensional theory, i.e. an extension of the previous study \cite{Ciupke:2015msa}, could in principle modify the strength of the $F^4$ corrections by affecting the values of the $\lambda$ parameter. Thus, in the case of a constant superpotential, one could argue that the scale of $\lambda$ parameter is expected to be enhanced $\lambda > \mathcal{O}(10^{-3})$. Our example could be seen as an example of the importance of these $F^4$ corrections in the moduli stabilization. One more crucial comment that needs to be highlighted is the fact that if the K{\"a}hler dependence of the those corrections is changed due to the extended operators, the form of the effective potential drastically changed. We refer for relevant discussion to \cite{Lacombe:2024jac}. In the following section, we continue our discussion on the fundamental concepts necessary for conducting a three-field inflationary analysis.

\section{Basic idea of multi-field inflation}\label{section3}

The purpose of this section is to set the stage for the  three-field inflationary scenario that we are going to present in the next section. We begin by writing the Lagrangian for multiple scalar fields minimally coupled with gravity in $(3+1)-$dimensional curved spacetime:
\begin{equation}
    S=\int d^4x \sqrt{-g}\left[M_{pl}^2\frac{R_4}{2}-\frac{\mathcal{G}_{ab}}{2}\partial_{\mu}\phi^a\partial^{\mu}\phi^b-V(\phi^a)\right],
\end{equation}
where Greek letters label the 4d spacetime indices of Friedman-Lemaitre-Robertson-Walker (FLRW) metric with $(-,+,+,+)$ metric signature. Lower latin indices denote field-space entries with $(a,b\equiv 1,2,3)$ and the metric is $\mathcal{G}_{ab}$. $V(\phi^a)$ is the scalar potential in terms of $\phi^a$. $M_{pl}^2=(8\pi G_N)^{-1}$ is the reduced Planck mass with $G_N$ being the Newton's constant. We will work in natural units, hence from now on we will put $M_{pl}=1$. In an FLRW spacetime, Einstein's equation of motion for a scalar field known as inflaton, takes the following form,
\begin{align}
    & H^2=\frac{1}{3}\left(\frac{\dot{\phi}^2}{2}+V(\phi^a)\right),\label{Hubble_constraint}\\
    & \ddot{\phi}^a+3H\dot{\phi}^a+\Gamma^a_{bc}\dot\phi^b\dot\phi^c+\mathcal{G}^{ab}V_b=0,\label{field_eom}
\end{align}
here
\begin{equation}
    \dot\phi^2\equiv \mathcal{G}_{ab}\dot\phi^a\dot\phi^b.
\end{equation}
Note that $H$ is the Hubble parameter overseeing the expansion of the spacetime during inflation and it is related to the scale factor of FLRW metric. The Christoffel symbols in \eqref{field_eom} is calculated from the field space metric $\mathcal{G}_{ab}$ and $V_a$ is the derivative of the potential with respect to the scalar field $\phi^a$. In multi-field inflationary models, \eqref{Hubble_constraint} acts as a constraint and it is enough to solve just the field equations. The equations above are written in physical time but for our purpose of studying inflation, it is more convenient to switch to number of efolds $(N)$ which is related to time as $dN=Hdt$.\par
Inflation is a nearly exponential evolution of the space-time. This is ensured by requiring the fractional change of the Hubble parameter per efold, denoted by $d\ln H/dN$ to be small. In the same spirit, one can define a series of Hubble slow-roll parameters $\epsilon_i$s where $\epsilon_1=-d\ln H/dN$. $\epsilon_1$ is the first Hubble slow-roll parameter and for brevity we will drop the subscript. In order for inflation to last at least $60$ efolds to, necessary to solve the horizon problem, $\epsilon$ should remain small for a sufficient number of Hubble times and this is measured by the second Hubble slow-roll parameter $\eta$, defined as:
\begin{equation}\label{hublle_slowroll}
    \eta\equiv \frac{\dot\epsilon}{\epsilon H}=\frac{\ddot H}{H\dot H}+2\epsilon=2\delta_{\phi}+2\epsilon\ll 1,
\end{equation}
where a small $\eta$ also requires a small $\delta_{\phi}$.

\subsection{Kinematic basis decomposition}
\begin{figure}[H]
 \centering
  \includegraphics[width=0.6\linewidth]{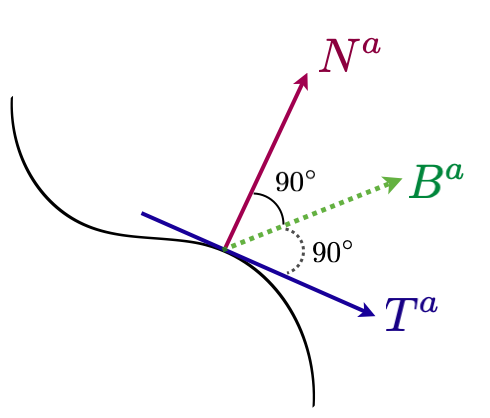}
  \caption{The black line corresponds to the inflationary trajectory on which $T^a$ is an unit vector tangential to it. $N^a$ is another unit vector normal to the trajectory and $B^a$ denotes the unit vector normal to both $T^a$ and $N^a$.}
\label{fig2}
\end{figure}

We are interested in a three-field inflation with turning and torsion in field space. Hence, it is natural to work on kinetic basis by decomposing the inflationary trajectory into tangent, normal and a binormal direction as shown in figure \ref{fig2}. An extensive study on three-field inflation called helix-trajectory inflation using an hyperbolic metric is carried out in \cite{Aragam:2019omo} and further extended in \cite{Christodoulidis:2022vww,Aragam:2023adu,Aragam:2024nej}. In this lore, let us introduce unit tangent $(T^a)$, normal $(N^a)$ and binormal $(B^a)$ vectors  as following:
\begin{align}
   & T^a=\frac{\dot\phi^a}{\dot\phi},\qquad N^a=\frac{\mathcal{D}_NT^a}{\Omega_1},\qquad B^a=\frac{D_N N^a+\Omega_1 T^a}{\tau},\nonumber\\
   & T^aT_a=B_aB^a=N_aN^a=1,\qquad N^aT_a=B^aT_a=N^aB_a=0,\nonumber\\
   & \Omega_1=\frac{\Omega}{H},\qquad \Omega=|\mathcal{D}_tT|,\qquad \tau B^a=\mathcal{D}_N N^a+\Omega_1 T^a.
\end{align}
$\Omega_1$ is known as the dimensionless \emph{turning rate}, $\Omega$ is dimensionful turning rate and $\tau$  denotes the dimensionless \emph{torsion} \cite{Pinol:2020kvw,Christodoulidis:2022vww}. Both of them determine how much an inflationary trajectory deviates from a geodesic one $(\mathcal{D}_tT=0,\,\,T=\sqrt{\mathcal{G}_{ab}T^aT^b})$. In a three-dimensional field-space $(x,y,z)$ for example, $\Omega$ denotes how much the inflationary trajectory is bending in the $x-y$ plane and $\tau$ denotes what is its motion along the vertical $z-$direction. When $\Omega>0$ the trajectory undergoes turning and when $\tau>0$ the trajectory is non-planar. If both $(\Omega,\tau)$ are constant, the trajectory takes a helical path as found in \cite{Aragam:2019omo}. Note that, the kinematic basis as well as the torsion is not well defined when $\Omega=0$. In our work however, as presented in section \ref{section 4}, the non-geodesic behavior will be manifested not in their kinetic terms but in the scalar potential through non-trivial coupling of scalar fields. $\Omega$ denotes the norm of the unit tangent vector and has one mass dimension. These can be summarised in terms of Frenet-Serret system as follows:
\begin{equation}
    \mathcal{D}_N\begin{pmatrix}
T^a \\
N^a \\
B^a 
\end{pmatrix}
= \begin{pmatrix}
0 & \Omega_1 & 0 \\
-\Omega_1 & 0 & \tau \\
0 & -\tau & 0 
\end{pmatrix}
\begin{pmatrix}
T^a \\
N^a \\
B^a 
\end{pmatrix}.
\end{equation}

With the help of the equation of motion \eqref{field_eom}, the definitions of $\Omega_1$ and $\tau$ can be related to the inflationary potential as \cite{Aragam:2023adu}:
\begin{align}
    \Omega &=-\frac{V_N}{H\phi'},\qquad '=\frac{d}{dN},\label{omega}\\
    \tau&=-\frac{V_{BT}}{\Omega_1 H^2},\label{torsion}
\end{align}
where $V_N=N^aV_a$, $V_{BT}=V_{ab}T^aB^b$ and $\nabla_aA_b\equiv \partial_aA_b-\Gamma^c_{ab}A_c$, for some vector $A_b$.

To study the masses of the scalar fields as well as to study multi-field perturbations, we define two matrices. $\mathbb{M}^a_b$ is the mass-matrix of the scalar fields computed from $V(\phi^a)$ and the projection of the $\mathbb{M}^a_b$ along the directions of the unit vectors is denoted by $\mathbb{M}$:
\begin{equation}\label{mass_matrix}
    \mathbb{M}^a_b=\nabla^a\nabla_bV,\qquad \mathbb{M}=\begin{pmatrix}
V_{TT} & V_{TN} & V_{TB} \\
V_{NT} & V_{NN} & V_{NB} \\
V_{BT} & V_{BN} & V_{BB} \\
\end{pmatrix},
\end{equation}
where $V_{TT}=T^aT^b\nabla_a\nabla_bV$.

\subsection{Multi-field perturbations}

In case of multi-field inflation, it is standard to decompose the linear equation of motion for Mukhanov-Sasaki variable in kinematic basis \ie along $(T^a,N^a,B^a)$. The projection of field fluctuations $Q^a$ along the unit vectors are $Q_i=(Q_T,Q_N,Q_B)\equiv (Q^aT_a,Q^aN_a,Q^aB_a)$. The equations of motion in terms of $Q_i$ are \cite{Pinol:2020kvw,Aragam:2023adu,Aragam:2024nej,Ashoorioon:2025iid}:
\begin{align}
   & \mathcal{D}_N(Q')^i+F_j^i(Q')^j+C_j^iQ^j=0,\nonumber\\
   & F^i_j\equiv (3-\epsilon)\delta^i_j-2\Omega^i_j,\nonumber\\
   & C_j^i=\left(\frac{k}{aH}\right)^2\delta^i_j+\begin{pmatrix}
0 & -2(3-\epsilon)\Omega_1 & 0 \\
0 & \mathcal{M}_{NN}-\Omega_1^2-\tau^2 & \mathcal{M}_{NB}-\tau(3-\epsilon) \\
0 & \mathcal{M}_{NB}+\tau(3-\epsilon) & \mathcal{M}_{BB}-\tau^2 
\end{pmatrix}+\mathcal{O}(\epsilon^2,\eta,\nu,\nu_{\tau}).\label{pertb_Cij}
\end{align}
Primes denote a derivative with respect to conformal time $d\tau=dt/a(t)$. We defined $\nu=\Omega_1'/\Omega_1$ and $\nu_{\tau}=\tau'/\tau$. The mass terms appearing in \eqref{pertb_Cij} such as $(\mathcal{M}_{NN},\mathcal{M}_{NB},\mathcal{M}_{BB})$ can be obtained by contracting $\mathcal{M}_{ab}$ with $(N^a,B^a)$ where $\mathcal{M}_{ab}$ is:
\begin{equation}\label{mass_mab}
    \mathcal{M}_{ab}=\frac{V_{ab}}{H^2}-2\epsilon R_{aTTb}+2\epsilon(3-\epsilon)T^aT^b+\sqrt{2\epsilon}\frac{T_aV_b+T_bV_a}{H^2}
\end{equation}
In order to further simplify the above set of equations one has to compare between the adiabatic mass and entropic masses. If the entropic masses are larger or comparable to the adiabatic mass, one has to consider the effect of all the fields in the perturbative analysis which thus modify  the spectral index and spectral tilt \cite{Aragam:2019omo} through a parameter called speed of sound $c_s$. It is the propagation speed of the adiabatic perturbation, expressed in terms of the ratio of entropic masses. The study of perturbation with reduced speed of sound and non-trivial turns is studied for the two-field case in \cite{Achucarro:2012sm,Achucarro:2012yr,Hetz:2016ics} and in the three-field scenarios in \cite{Kaiser:2012ak,Cespedes:2013rda}. On the other hand, for the inflationary analysis done in this paper --- the mass gap between the adiabatic and entropic masses are large enough so that we integrate out the latter. In this case, we can study the inflationary perturbation as a quasi-single field effective theory, knowing that it is subject to $M_{TT}/M_{\text{entropic}}$ corrections.

\section{Three-field analysis}\label{section 4}

In this section, we shall thoroughly analyze the complete multi-field inflationary evolution and the detailed predictions of the model. To achieve this, we begin by outlining the consistency checkpoints that must be met in a viable string theory model building.

\subsubsection*{Consistency checks}

\begin{enumerate}
    \item To reside in the semi-classical regime, we need to achieve small $g_s$ and large volume $\mathcal{V}$. From the examples presented in table (\ref{tab1},\ref{tab2}), we can observe that all of them meet the specified criteria with $g_s\sim\mathcal{O}(10^{-4})$ and $\mathcal{V}\sim\mathcal{O}(10^6)$. Due to the same reason, it is easier to trust the underlying EFT in our model by requiring:
    \begin{equation*}
        V_{\alpha'^3}>V_{F^4},
    \end{equation*}
    which means the leading order BBHL-term proportional to $\hat{\xi}$ in the perturbative expansion in $\alpha'$ should be larger compared to the higher derivative $F^4$ term, proportional to $\lambda$. The above inequality simplifies to:
    \begin{equation}\label{cond1}
        \mathcal{V}>\frac{W_0^2\lambda}{2\pi^2\sqrt{g_s}\hat\xi},
    \end{equation}
    where $\mathcal{V}$ is the volume in \eqref{volume2}. For a typical set of parameters, one can show that the inequality poses a less strict constraint for the parameter sets presented in this paper. For the inflationary example as well as the models presented in table (\ref{tab1},\ref{tab2}), we get $\mathcal{V}>\mathcal{O}(10^2)\lambda$, with $\lambda\ll 1$. 
    \item High energy stringy effects can be neglected if  each four-cycle considered in this work, which are $\tau_1,\tau_6,\tau_7$, has volume larger than the string scale\footnote{For details on how the volumes and 10D metric in Einstein and string frame are related see \cite{ValeixoBento:2023afn}.}: $\text{Vol}_s^{1/4}\gg \sqrt{\alpha'}$. The string frame and Einstein frame volume is related as: $\text{Vol}_s=g_s\text{Vol}_E=g_s\tau_E l_s$ with $l_s=2\pi \sqrt{\alpha'}$. We arrive at the following relation:
\begin{equation}\label{cond2}
    \delta_{\tau_i}\equiv \frac{1}{g_s(2\pi)^4\tau_i}\ll 1\qquad \forall i.
\end{equation}
From table-\ref{tab1}, we see that with $\tau_i\sim \mathcal{O}(10^3-10^4)$ and $g_s\sim\mathcal{O}(10^{-4})$, it is straightforward to satisfy this condition.

\item To trust the $\alpha'$ expansion, one has to satisfy:
$\delta_{\xi}=\frac{\hat\xi}{2\mathcal{V}}\ll 1$. For $\hat\xi\sim\mathcal{O}(10)$ and $\mathcal{V}\sim\mathcal{O}(10^6)$, it is easily satisfied in our models.
\item Again from the consistency of the EFT, throughout the inflationary evolution, we need\footnote{See \cite{Chakraborty:2019dfh} for a slightly different hierarchy of scales where the inequalities between inflaton mass $m_{\text{inf}}$ and the underlying Hubble scale is interchanged to $m_{\text{inf}}>H$ throughout the slow-roll inflationary evolution in a multi-field scenario.}:
\begin{equation}\label{inf_hierarchy}
    m_{\varphi_i}<H<M_{kk}^{\text{bulk}}<M_s<M_{pl}.
\end{equation}
\item Last but not least, the inflationary prediction should be in compliance with the Planck 2018 data \cite{Planck:2018jri}:
\begin{equation}\label{planck_bound}
    P_s=2.105\pm 0.03\times 10^{-9},\qquad n_s=0.9649\pm 0.0042,\qquad r<0.036, 
\end{equation}
    at a pivot scale of $0.05\,\mathrm{Mpc^{-1}}$. 
\end{enumerate}

\subsection{Cosmological evolution}

We now analyze the three-field cosmological evolution in a fully canonical basis spanned by $(\varphi_1,\varphi_2,\varphi_3)$. The field space metric being flat makes both the Christoffel symbols and the Ricci scalar zero. In this setup, the evolution equations of \eqref{field_eom}, in the canonical basis boils down to these simpler set of equations in number of efolds:
\begin{align}\label{cano_eom}
    \varphi_1''&=-\left(\frac{1}{2}\left(\varphi_1'^2+\varphi_2'^2+\varphi_3'^2\right)+3\right)\varphi_1'-\frac{V_{,\varphi_1}}{H^2},\nonumber\\
        \varphi_2''&=-\left(\frac{1}{2}\left(\varphi_1'^2+\varphi_2'^2+\varphi_3'^2\right)+3\right)\varphi_2'-\frac{V_{,\varphi_2}}{H^2},\nonumber\\
            \varphi_3''&=-\left(\frac{1}{2}\left(\varphi_1'^2+\varphi_2'^2+\varphi_3'^2\right)+3\right)\varphi_3'-\frac{V_{,\varphi_3}}{H^2},\qquad '=\frac{d}{dN}.
\end{align}
%write down the eoms for your models and show the plots one by one plots to add: three fields, epsilon,eta, omega, torsion, mass scales
$V$ is the scalar potential in \eqref{exact} which consists of leading order $\alpha'^3$ BBHL-term, log-loop corrections, higher derivative $F^4$ corrections and a constant uplift term resulting to a Minkowski-like vacuum of the potential at its global minimum. We are agnostic about the exact computation of the uplift term, it can either be a D-term uplift \cite{Burgess:2003ic}, a T-brane uplift \cite{Cicoli:2015ylx} or an anti-brane uplift \cite{Kachru:2002sk,Crino:2020qwk,Vafa:2005ui,Cicoli:2024bxw}. \par
We now present a three-field inflationary model where all the three fields $(\varphi_1,\varphi_2,\varphi_3)$ are displaced from their respective minimum. We get sufficient efolds as these fields complete their journey towards their minima and oscillate there to start the reheating process\footnote{Note that, we do not intend to study the reheating dynamics in this article}. Below, we discuss one benchmark model which uses the following set of parameters:
\begin{table}[H]
\begin{center}
\centering
%\begin{adjustbox}{width=0.9\textwidth}
\begin{tabular}{| l | c | c | c | c | c | c | c | c |c |c |c |c|c|c |l |}
\hline
\cellcolor[gray]{0.9} $g_s$ &\cellcolor[gray]{0.9} $|W_0|$ &  \cellcolor[gray]{0.9} $|\hat{\eta}|$ & \cellcolor[gray]{0.9} $\alpha$  & \cellcolor[gray]{0.9} $\hat\xi$ & \cellcolor[gray]{0.9} $\lambda$ & \cellcolor[gray]{0.9} $C_{\text{up}}$ & \cellcolor[gray]{0.9} $\varphi_1^0$ & \cellcolor[gray]{0.9} $\varphi_2^0$ & \cellcolor[gray]{0.9} $\varphi_3^0$ & \cellcolor[gray]{0.9} $\langle\varphi_1\rangle$ &\cellcolor[gray]{0.9} $\langle\varphi_2\rangle$ &  \cellcolor[gray]{0.9} $\langle\varphi_3\rangle$  \\
\hline \hline
 $0.001$ & $38$ &  $0.5$ & $0.0001$ & $16$ & $0.008$ & $2.45\times 10^{-14}$ & $4.9$ & $10.0$ & $1.1$ & $7.09$ & $9.06$ &  $1.15437$  \\
\hline
\end{tabular}
%\end{adjustbox}
\end{center} 
\caption {The table contains the model parameters and initial conditions. The latter is labelled by superscript $0$.}
\label{tab3}
\end{table}
\begin{table}[H]
\begin{center}
\centering
%\begin{adjustbox}{width=0.9\textwidth}
\begin{tabular}{| l  | c | c | c | c | c |c |c |c |c |l |}
\hline
 \cellcolor[gray]{0.9} $m_{\langle\varphi_1\rangle}^2$  & \cellcolor[gray]{0.9} $m_{\langle\varphi_2\rangle}^2$ & \cellcolor[gray]{0.9} $m_{\langle\varphi_3\rangle}^2$ & \cellcolor[gray]{0.9} $M_s/M_{pl}$ & \cellcolor[gray]{0.9} $M_{KK}/M_{pl}$ & \cellcolor[gray]{0.9} $m_{3/2}/M_{pl}$ & \cellcolor[gray]{0.9} $H/M_{pl}$  \\
\hline \hline
   $1.54\times 10^{-6}$ & $2.51\times 10^{-14}$ & $5.33\times 10^{-15}$ & $1.36\times 10^{-3}$ & $1.25\times 10^{-3}$ & $4.5\times 10^{-6}$ & $\sim 6.95\times 10^{-8}$ \\
\hline
\end{tabular}
%\end{adjustbox}
\end{center} 
\caption {Mass scales follow the correct hierarchy. The Hubble parameter, derived from \eqref{Hubble_constraint}, remains nearly constant throughout the inflationary evolution, which is why a wiggle sign has been included. }
\label{tab:4}
\end{table}
Masses of the inflatons tabulated above are Eigenvalues of the Hessian matrix computed using \eqref{exact} at their respective minimum. In accordance with the consistency checks (\eqref{hierarchy1} and \eqref{inf_hierarchy}) stated above, the hierarchy of various mass scales are satisfied \ie:
\begin{equation*}
    m_{3/2}<M_{KK}<M_s<M_{pl},\qquad \text{and}\qquad M_{\varphi_i}\equiv M_{\text{inf}}<H.
\end{equation*}
For this benchmark model, the overall volume of \eqref{volume2} is $\mathcal{V}\sim \mathcal{O}(10^4)$. Residing in the large volume and the weak coupling limit, it is straightforward for this model to satisfy condition \eqref{cond1} and we obtain $\delta_{\xi}\sim \mathcal{O}(10^{-4})$ and $\delta_{\tau_i}\sim \mathcal{O}(10^{-5})$.

\subsubsection*{Slow-roll dynamics of the fields}

We numerically evolve the set of equations presented in \eqref{cano_eom} with the values of the model parameters and initial conditions presented in table \ref{tab3}. We are focusing on vanishing initial velocities for the inflatons \ie $\{\varphi_1'(0),\varphi_2'(0),\varphi_3'(0)\}=0$.
\begin{figure}[ht]
    \centering
    % Subfigure 1
    \begin{subfigure}[b]{0.45\textwidth}
        \centering
        \includegraphics[width=\textwidth]{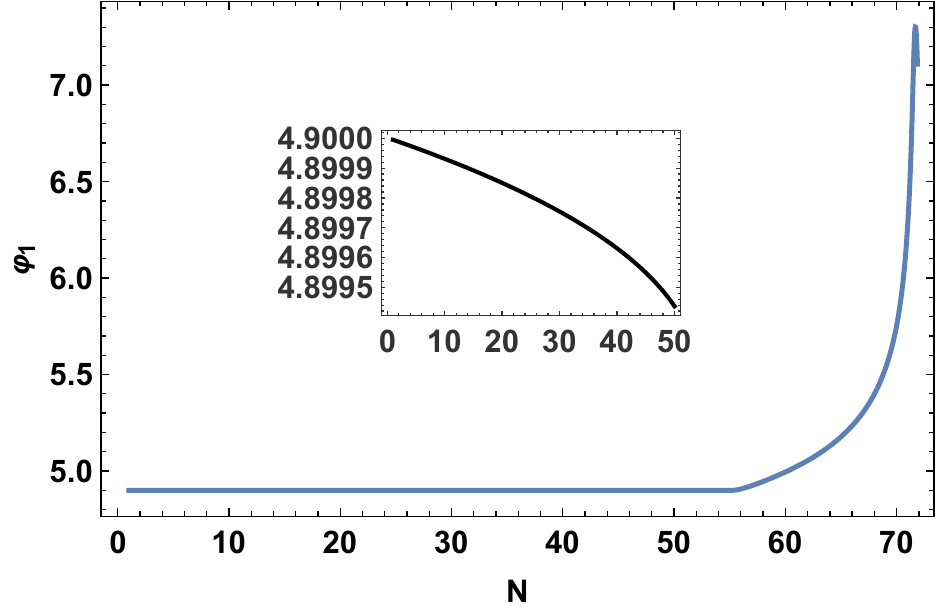}
       % \caption{(a))}
        \label{fig:phi1}
    \end{subfigure}
    \hfill
    % Subfigure 2
    \begin{subfigure}[b]{0.45\textwidth}
        \centering
        \includegraphics[width=\textwidth]{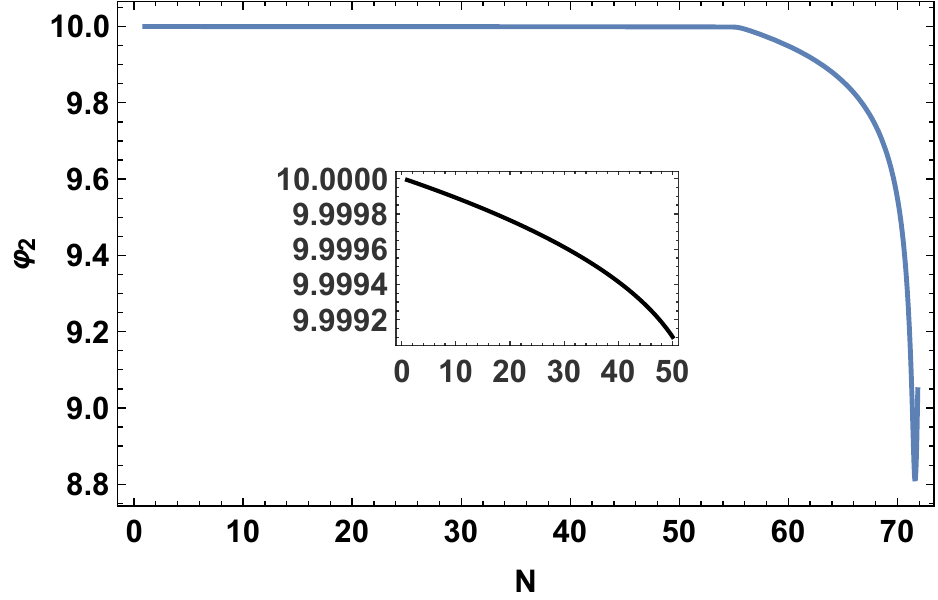}
       % \caption{}
        \label{fig:phi2}
    \end{subfigure}
    %\caption{Overall caption for the figure.}
 %   \label{fig:main}
\end{figure}
\begin{figure}[ht]
    \centering
    % Subfigure 1
    \begin{subfigure}[b]{0.45\textwidth}
        \centering
        \includegraphics[width=\textwidth]{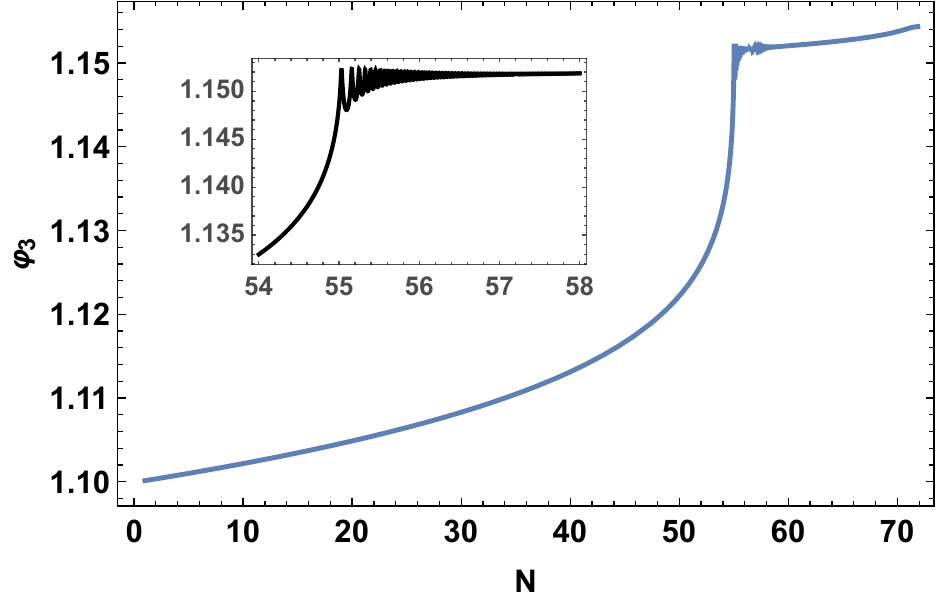}
        %\caption{}
        \label{fig:phi3}
    \end{subfigure}
    \hfill
    % Subfigure 2
    \begin{subfigure}[b]{0.5\textwidth}
        \centering
        \includegraphics[width=\textwidth]{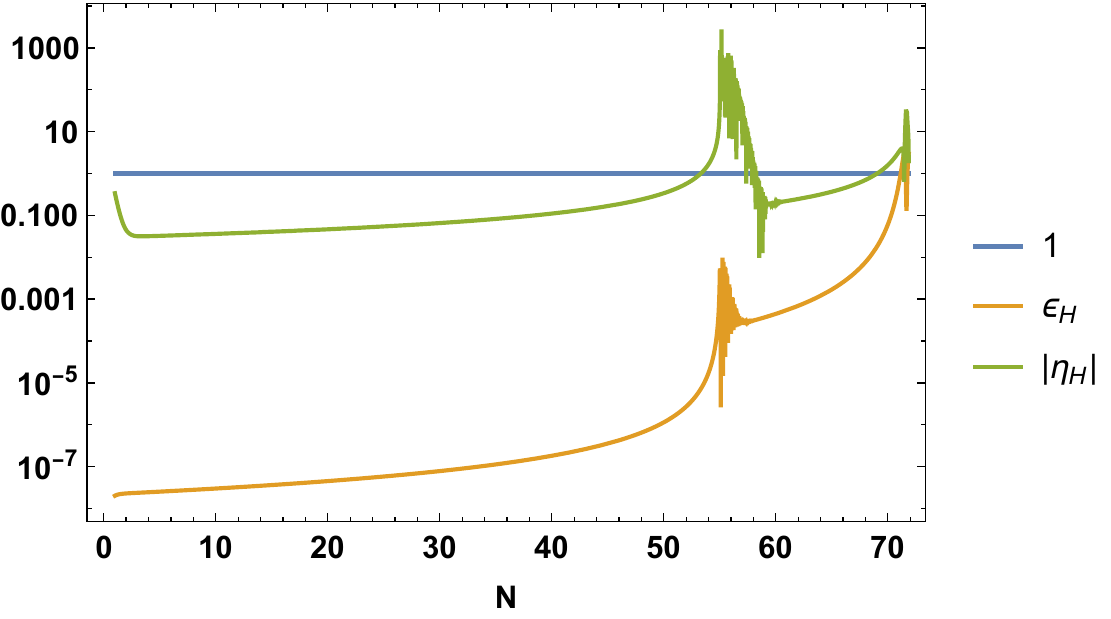}
       % \caption{}
        \label{fig:epseta}
    \end{subfigure}
    \caption{Field evolutions of the three canonical scalar fields. $\epsilon_H$ stays less than $1$ throughout the evolution but $\eta_H$ crosses $1$ at $N=55$. The reason for the same have been explained in the text.}
    \label{fig:4}
\end{figure}

If we assess the initial conditions for the inflatons and their respective minima in table \ref{tab3}, we see that the initial path of the inflationary trajectory starts on the left side of $\langle\varphi_1\rangle$ and $\langle\varphi_2\rangle$. While both $(\varphi_1,\varphi_2)$ are displaced  away from their minima, if we choose $\varphi_3$ close to its minimum, we obtain an interesting three-field attarctor where all three-fields move together to give sufficient amount of efolds. As soon as the fields settle at their minima, inflation ends. In Figure \ref{fig:4}, although along $\varphi_1$ and $\varphi_2$ the displacement in fields $(\Delta\varphi_{1,2})$ remain relatively small in most of their evolution, they do exhibit some movement, as evident from the inset images. The potential along the $\varphi_3$ is the steepest among the rest, it dominates the field trajectory in the beginning when the $(\varphi_1,\varphi_2)$ move relatively slower. Hence, we may conclude that most of the slow-roll dynamics upto $N=55$ is mostly a cumulative effect of these fields with contributions from $\varphi_3$ being the dominant one. An interesting observation here is that, although Table \ref{tab:4} shows that \(\varphi_3\) has the lowest mass at the global minimum, as the inflaton evolves from a shifted position, this hierarchy reverses, making \(\varphi_3\) direction the steepest among all. \par

The trajectory of $\varphi_3$ shows a piculiar behavior because if we notice and compare between the path $\varphi_3$ follows and its value of the minimum from table \ref{tab3} then we can see that the field is not directly going towards its minimum. Roughly $N=55$ onwards, we notice vigorous oscillations in both $\epsilon_H,\eta_H$ followed by a softening where $\eta_H$ returns to slow-roll after a brief period of slow-roll violation. We have $\epsilon_H<1$ throughout but its amplitude increases beyond $N=55$. The oscillation in $(\epsilon_H,\eta_H)$ is a manifestation of the oscillation in $\varphi_3$ direction around $N\sim 55$. \par

\begin{figure}[H]
        \centering
\includegraphics[width=0.6\textwidth]{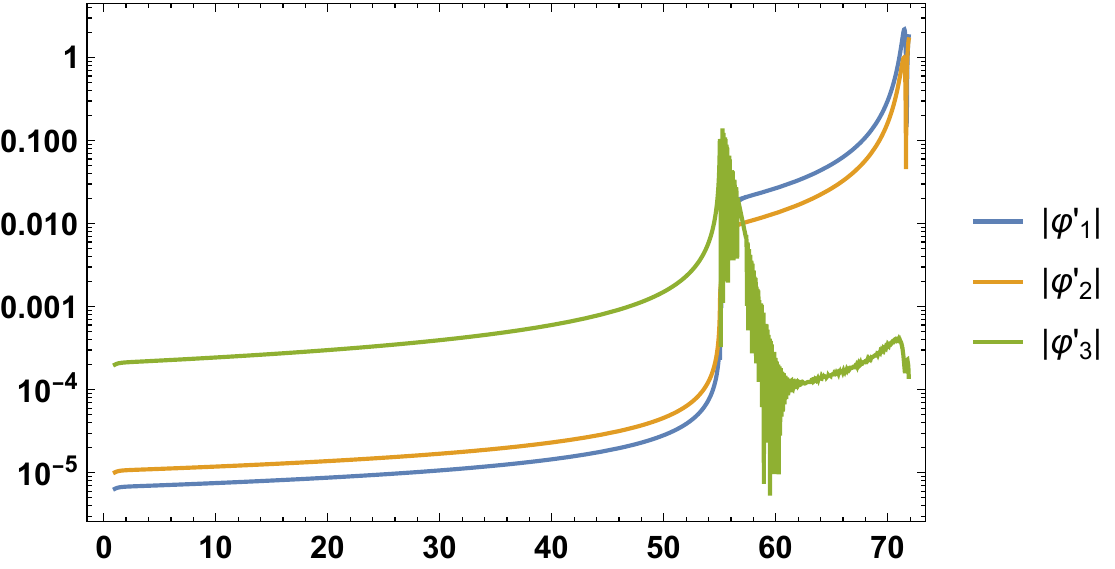}        
    \caption{Speed of the inflatons.}
    \label{fig5}
\end{figure}

We can understand this behavior by looking at the field trajectories as well as the steepness of the potential along $\varphi_3$. The reason is $\varphi_3$ falls in a ridge of the potential. We notice that along the inflationary trajectory where $(\varphi_1,\varphi_2)$ are far away from their minima, the true minimum for $\varphi_3$ disappears and the ridge becomes unavoidable. Although, notice that the ridge is not at the point where the potential diverges which is at $\varphi_3\equiv \sqrt{4/3}$. The ridge acts like a false vacuum for $\varphi_3$ and when it falls there it leads to such oscillations\footnote{Note that, it may be possible that this oscillation along $\varphi_3$ may trigger an early preheating. However, since slow-roll resumes after the end of oscillation, leading to the continued expansion of spacetime, any particles generated during this period will ultimately be diluted away.}. Now, as soon as $(\varphi_1,\varphi_2)$ move close to their minima along the trajectory, $\varphi_3$ recovers its actual minimum and $\varphi_3$ restarts its journey towards the true vacuum. Interestingly, it appears that after the oscillation the fields switch their roles, $(\varphi_1,\varphi_2)$ move faster whereas $\varphi_3$ appears relatively slower, see figure \ref{fig5}. For the last few efolds, all the fields become active to roll towards the global minimum, thereby increasing both $\epsilon_H$ and $\eta_H$, signalling the end of inflation. 
As expected, $\varphi_3$ being the steepest among them, its velocity dominates over the other and as soon as it falls off the ridge the velocity decreases but does not become zero. When $\varphi_3$ oscillates at the ridge, $(\varphi_1,\varphi_2)$ become dominant and move towards their minimum. In summary, this attarctor solution we have discovered, gives a \emph{quasi-single-field} behavior in the first part of inflation and \emph{quasi-double-field} for the second period of inflation. We shall confirm this claim by looking at the projection of the masses of the scalar fields as well as perturbative masses of the modes throughout the inflationary evolution. We shall discuss this in greater detail in the following subsections.\par
To complete the analysis of this section, we also confirm that our benchmark model produces the correct hierarchy of scales as shown below.
\begin{figure}[H]
        \centering
\includegraphics[width=0.6\textwidth]{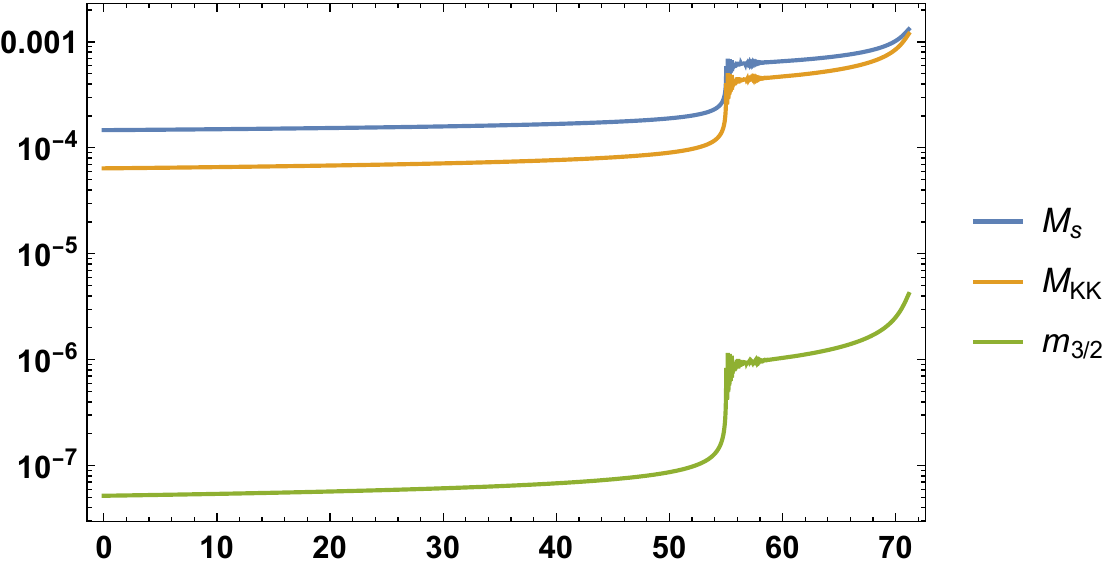}        \label{fig6}
    \caption{Correct hierarchy of mass scales with $M_s>M_{KK}>m_{3/2}$. All the masses are expressed in Planck units.}
\end{figure}

\subsubsection*{Turning rate and torsion}

\begin{figure}[H]
        \centering
\includegraphics[width=0.6\textwidth]{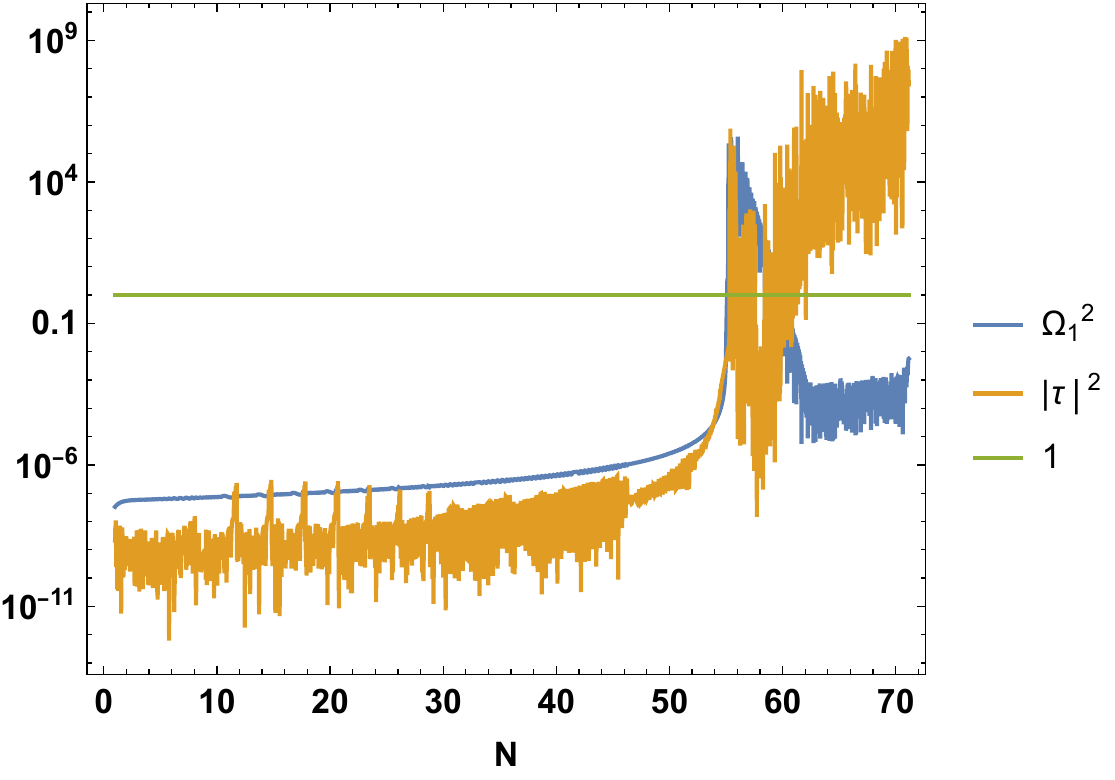}      
    \caption{The dimensionless turning rate and torsion}
      \label{fig7}
\end{figure}

Depending on the nature of turns and torsions, one can get a variety of different models with different predictions for cosmological observables. Models featuring sharp turns \cite{Parra:2024usv}, smooth and large turns \cite{Ishikawa:2021xya}, and turns combined with non-planar motion \cite{Aragam:2019omo,Aragam:2024nej} have proven useful in the study of primordial black holes and particle production. For our multi-field canonical model with vanishing Ricci curvature, it is expected that the fields follow a geodesic $(\mathcal{D}_tT=0)$ for most of its evolution producing small $\Omega/H\sim \mathcal{O}(10^{-8})$ and $\tau\sim\mathcal{O}(10^{-12})$. During the crossing of pivot-scale $(k_{\star}=0.05\,\mathrm{Mpc^{-1}})$ which roughly corresponds to a scale between $(N_{\epsilon_H=1}-60)=11.15$ to $(N_{\epsilon_H=1}-50)=21.15$, $(\Omega/H,\tau)$ remain small, however if we look at figure \ref{fig7}, we notice that beyond $N=55$, both $(\Omega/H,\tau)$ get an enhancement in magnitude. The rapid oscillation observed in the figure below arises from the oscillation in the $\varphi_3$ direction as it falls into its false vacuum (the ridge), as we can observe in figure \ref{fig:4}. 

The oscillations are present in both $\Omega/H$ and $\tau$ direction but notice that post-oscillation the hierarchy between the torsion and the turning rate changes. This signals the fact that, the field-motion was confined within a geodesic trajectory dominated by the $\varphi_3$ field but as soon as $(\varphi_1,\varphi_2)$ find the valley towards its minima and start their journey, the whole inflationary trajectory becomes non-planar. 

\subsection{Cosmological parameters}

The dynamics of the linear perturbation and cosmological predictions for an inflationary model depend on the hierarchies of the adiabatic and the entropic modes. The value of $(\Omega/H,\tau)$ play a pivotal role in understanding how much entropic perturbations affect the adiabatic curvature perturbation. As we can see from the set of equations \eqref{pertb_Cij}, $\tau\sim 0$ will simplify them to the standard Mukhanov-Sasaki mode equations in terms of two canonically normalised fields $v_{\sigma}$ and $v_s$ which depend on $(Q_T,Q_N)$ in the following way:
\begin{equation}
    v_{\sigma}=aQ_{\sigma},\qquad v_s=aQ_N.
\end{equation}
In terms of these new variables, the mode equations in conformal time can be written as:
\begin{align}
 &   v_{\sigma}''-\xi' v_s'+\left(k^2-\frac{z''}{z}\right)v_{\sigma}-\frac{(z\xi')'}{z}v_s=0,\\
 & v_{s}''+\xi' v_{\sigma}'+\left(k^2-\frac{a''}{a}+a^2\mu_s^2\right)v_{s}-\frac{z'}{z}\xi' v_{\sigma}=0,\qquad z=a\dot\varphi/H.
\end{align}
$\xi'=2a\Omega$ sources the coupling between curvature and entropic mode. $\Omega\ll 1$ naturally makes the coupling weak in our case. The masses of adiabatic and entropic perturbations can be derived from both \eqref{mass_mab} and the matrix \eqref{mass_matrix}. These quantities play a crucial role in determining the extent to which entropic perturbations influence the whole inflationary perturbations, guiding us either deeper into a multi-field regime or away from it. The masses are defined as follows:
\begin{align}
\frac{m_T^2}{H^2}&\equiv -\frac{3}{2}\eta-\frac{1}{4}\eta^2+\frac{1}{2}\epsilon \eta-\frac{1}{2}\frac{\dot{\eta}}{H},\qquad ^.=\frac{d}{dt},\\
\frac{M^2}{H^2}&=\frac{V_{NN}}{H^2}+M_{pl}^2\epsilon \mathbb{R}-\Omega_1^2,\\
\mathbb{R}^a_{bcd}&=\Gamma^a_{bd,c}-\Gamma^a_{bc,d}+\Gamma^a_{ce}\Gamma^e_{bd}-\Gamma^a_{de}\Gamma^e_{bc},
\end{align}
Below, we show a plot of $(m_T^2,M^2)$ together with the mass along the binormal unit vector $\mathcal{M}_{BB}^2$ and an effective entropy mass given as $\mathcal{M}_{eff}^2=M^2+4\Omega^2$:
\begin{figure}[H]
    \centering
    % Subfigure 1
    \begin{subfigure}[b]{0.48\textwidth}
        \centering
        \includegraphics[width=\textwidth]{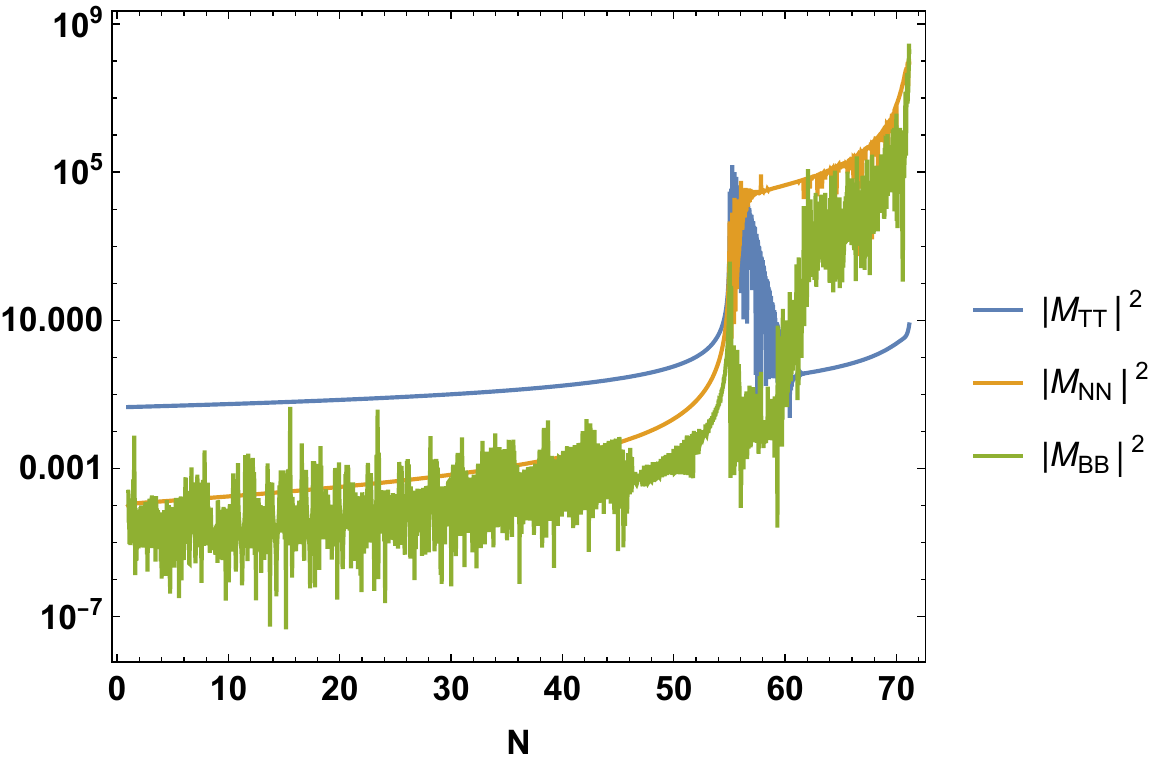}
        %\caption{}
        \label{fig:phi3}
    \end{subfigure}
    \hfill
    % Subfigure 2
    \begin{subfigure}[b]{0.48\textwidth}
        \centering
        \includegraphics[width=\textwidth]{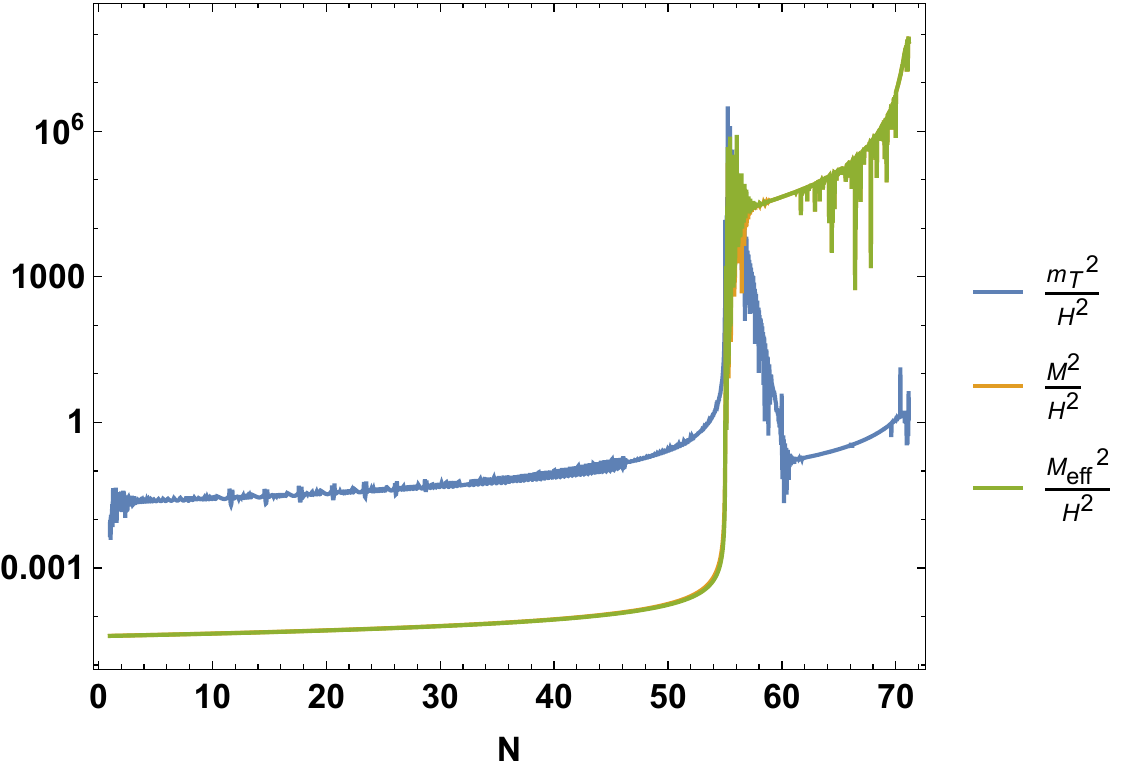}
       % \caption{}
    \end{subfigure}
    \caption{\textbf{Left:} the adiabatic mass is bigger than the entropic ones: $M_{TT}>M_{NN}>M_{BB}$. \textbf{Right:} similar hierarchy of scales are restored when the torsion is negligible. Also note that, we obtain $\mathcal{M}/\mathcal{M}_{eff}\sim 1$.}
            \label{fig:pertb_mass}
\end{figure}
The masses also help to define a speed of sound for the adiabatic perturbations via the relation \cite{Achucarro:2010da,Achucarro:2012yr,Cespedes:2012hu}:
\begin{equation*}
    c_s^{-2}=\frac{\mathcal{M}_{eff}^2}{M^2}.
\end{equation*}
We find that $\mathcal{M}_{\text{eff}} \sim M$, leading to $c_s \sim 1$ along the entire inflationary trajectory. Consequently, this results in negligible equilateral-type non-Gaussianity, with $ f_{NL} \sim \mathcal{O}(10^{-8})$, as given by the relation \cite{Aragam:2019omo}:
\begin{equation}
    f_{NL}=\frac{125}{108}\frac{\epsilon_H}{c_s^2}+\frac{5}{81}\frac{c_s^2}{2}\left(1-\frac{1}{c_s^2}\right)^2+\frac{35}{108}\left(1-\frac{1}{c_s^2}\right).
\end{equation}
After carefully assessing the mass-scales, we also move on to checking the adiabaticity condition \cite{Cespedes:2012hu}:
\begin{equation}
    \mathcal{A}=\left|\frac{\dot\Omega}{M\Omega}\right|\ll1,
\end{equation}
which holds true in our case with $\mathcal{A}\sim \mathcal{O}(10^{-5})$. Therefore, given the small value of $\mathcal{A}$ as well as the mass of the adiabatic perturbation being the dominant one, we use the standard formulae for the spectral index and spectral tilt in terms of the slow-roll parameters \eqref{hublle_slowroll}:
\begin{equation}
    n_s=1-2\epsilon_H-\eta_H,\qquad r=16\epsilon_H.
\end{equation}
The amplitude of the scalar power spectrum $(\Delta_s^2)$ and tensor power spectrum $(\Delta_t^2)$ are given by:
\begin{equation}
    \Delta_s^2=\frac{1}{8\pi^2}\frac{H^2}{\epsilon},\qquad \Delta_t^2=\frac{2H^2}{\pi^2}.
\end{equation}

%\textbf{DC: why r is small.$\to$ probably because V is small in $M_{pl}$ units.}

\begin{figure}[H]
    \centering
    % Subfigure 1
    \begin{subfigure}[b]{0.48\textwidth}
        \centering
        \includegraphics[width=\textwidth]{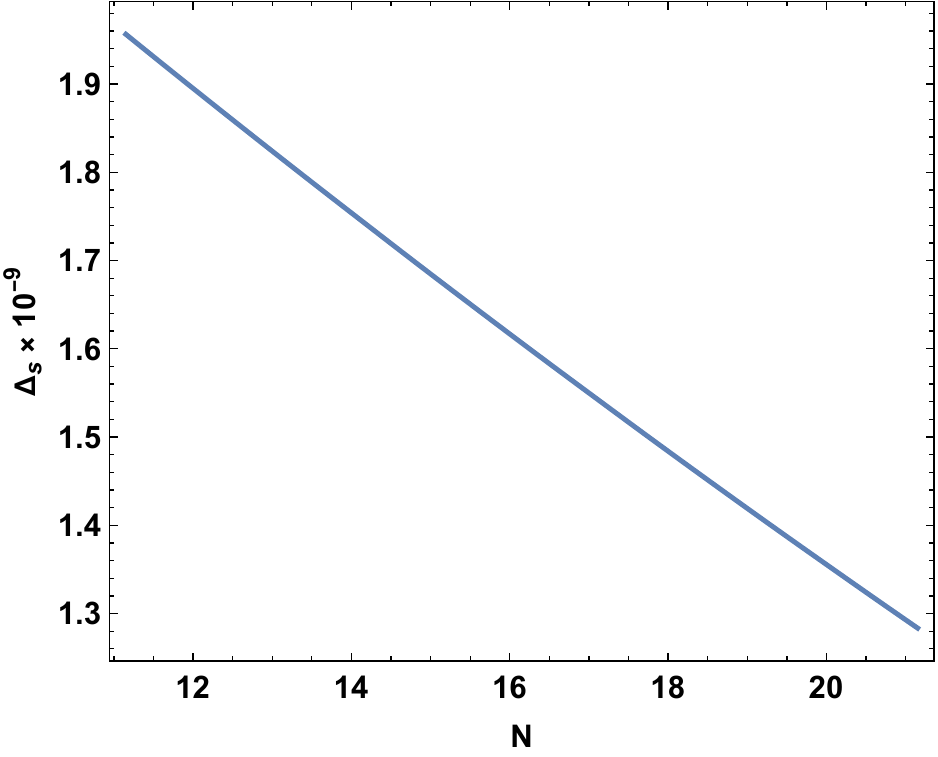}
        %\caption{}
        \label{fig:phi3}
    \end{subfigure}
    \hfill
    % Subfigure 2
    \begin{subfigure}[b]{0.48\textwidth}
        \centering
        \includegraphics[width=\textwidth]{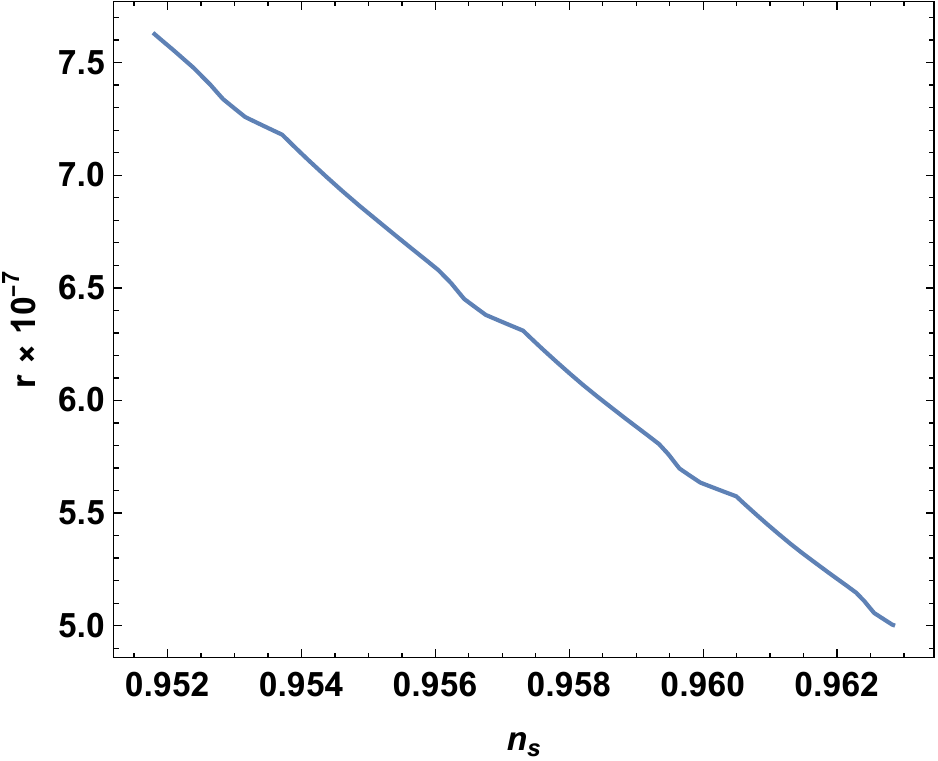}
       % \caption{}
        \label{fig:epseta}
    \end{subfigure}
    \caption{Scalar power spectrum $(\Delta_s^2)$, spectral index $(n_s)$ and spectral tilt $(r)$  are calculated at horizon crossing, that is roughly $50-60$ efolds before the end of inflation which is at $N_{\epsilon_H=1}=71.15$.}
    \label{fig:nsr_power}
\end{figure}
The figure above confirms that we are fairly within the Planck's bound presented in \eqref{planck_bound}. The small value of $r$ is a manifestation of the smallness of the scale of the potential in Planck units. Since $r$ is directly related to the field excursion through Lyth bound, in our case, we also expect small field excursions. This makes our model in accordance with the swampland distance conjecture \cite{Ooguri:2018wrx}. The running of the spectral index for this model, $\alpha_s$ is of the order $\mathcal{O}(10^{-3})$. 

\subsubsection*{Large turns and torsions on small scales}

Before ending this section, let us drop our final remark on the implications of the enhancement in the magnitude of turning rate and torsion as has been shown in figure \ref{fig7}. It is widely known in the literature \cite{Braglia:2020eai,Palma:2020ejf,Fumagalli:2020adf,Fumagalli:2020nvq,Ishikawa:2021xya,Fumagalli:2021mpc,Parra:2024usv} that models for which second slow-roll paramaters \ie $\eta_H$ is larger than $\epsilon_H$, together with a brief period of slow-roll violation is present -- the cumulative effect can potentially give rise to efficient production of primordial black holes. The field behavior reflected in our model reveals a similar behavior like double-inflation. In double inflation, the process occurs in two stages. Initially, one field rolls down to its minimum while the second field remains frozen. In the later stage, the frozen field begins to roll, ultimately reaching its minimum and bringing inflation to an end. Similarly, in the case of the three-field inflation discussed in this paper, we observe a \emph{quasi-double inflation}, where the first phase of inflation is mostly driven by one field, while the second phase is governed mostly by the other two fields as we can observe in figure \ref{fig5}. First phase of slow-roll takes place along the steep direction of the potential. \par

Referring to figure \ref{fig:pertb_mass}, we observe that during horizon crossing $(11.15<N<21.15)$, the mass of the adiabatic mode, denoted as $\{M_{TT}^2, m_T^2\}$, remains dominant. But on a smaller scale however, when the inflationary trajectory enters its second stage, characterized by two-field dynamics, where $(\varphi_1, \varphi_2)$ becomes more active compared to $\varphi_3$ -- the masses of the entropic perturbations become dominant. This implies that the effects of turns and torsions must be accounted for, requiring a simultaneous analysis of both adiabatic and all entropic perturbations to determine the power spectrum of scalar and tensor modes.\par %This highlights one of the most significant aspects of our paper. \par

Primordial black holes, considered potential progenitors of dark matter, have been widely studied in recent years, including a recent investigation within a three-field model in \cite{Aragam:2024nej} where the authors presented a toy model. We emphasize that our work is \emph{among the first to explore a three-field inflation scenario with non-trivial features at small scales ---  motivated by type-IIB string theory flux compactification where our inflatons are four-cycles of a K3-fibred Calabi-Yau.} The potential is a combination of leading order BBHL term, higher derivative $F^4$ term, and 1-loop correction terms. To ensure a nearly Minkowskian global minimum, a constant term has also been added. The fields are canonical and dimensionless in Planck units. We have performed a thorough consistency check to ensure the reliability of the 4d EFT. We plan to revisit the detailed study of primordial black holes as well as secondary induced gravitational waves  in a forthcoming companion paper.

\section{Conclusions}\label{section5}

In this paper, we have presented a three-field inflationary construction in string theory flux compactification for the first time where the hierarhcy between first and second slow-roll parameters with a transient violation of slow-roll becomes evident. The model is consistent with the Planck data at the horizon crossing. The scalar potential deviates from the usual pLVS \cite{Antoniadis:2018hqy} or fibre inflationary setup in their interplay of subleading corrections. The analysis is focused on the implications of logarithmic loop corrections and higher derivative $F^4$ correction in the moduli stabilization procedure. We also added a small uplifting term to ensure the presence of a Minkowski vacuum. \par

The volume of the CY is stabilized to a large value of order $\mathcal{V}>\mathcal{O}(10^6)$ (see table \ref{tab1}) and hence we can trust the underlying 4D low energy EFT in making the correction terms subleading in $\alpha'$ and $g_s$ expansions. The model reflects the correct hierarchy of mass scales, \ie $M_{pl}>M_s>M_{KK}>M_{inf}$ in all the benchmark models displayed in table \ref{tab1}. As far as the inflationary setup is concerned, we present a multi-field inflation model where all the scalar fields are active during inflation. Inflation occurs along $(\varphi_1,\varphi_2,\varphi_3)$ directions which span the canonical field space metric, constructed out of several four-cycles such as fibre modulus $\tau_7$, the volume of the base $\tau_6$ and the blow-up modulus $\tau_1$. Inflation starts at a position where all the scalars are shifted from their global minimum. However, it features a ridge-valley along the $\varphi_3$ direction when the other fields are away from their minimum. The potential in that direction becomes steeper than in the other two directions $(\varphi_1,\varphi_2)$. %This behavior results in a two-stage inflationary process. During the first stage, \(\varphi_3\)  rolls first, with \((\varphi_1, \varphi_2)\) shifting to a new position as the minimum for \(\varphi_3\) vanishes.\par

After analyzing the field trajectory, we confirm that inflation along this attractor occurs in two distinct stages. The first stage is mostly dominated by $\varphi_3$ because the potential in this direction is the steepest. The minimum along $\varphi_3$ disappears at $\{\varphi_1^{\text{shifted}},\varphi_2^{\text{shifted}}\}$, but forms a ridge ---compelling $\varphi_3$ to settle into it and oscillate, effectively behaving as a false vacuum for $\varphi_3$. The ridge continues to stay until the other two fields start to roll near their  respective minima --- signalling the onset of the second stage of inflation. When comparing the speed of the inflatons, the inflationary trajectory in the first stage is largely dominated by $\varphi_3$--- leading to a behavior of \emph{quasi-single field}. The horizon crossing of the pivot scale $(k_{\star})$ occurs in the first stage and the model is observationally viable. \par

In the second stage, as $\varphi_3$ stops its oscillation, the other two fields start their journey towards their minima. As soon as $(\varphi_1,\varphi_2)$ start to get near their minima, the true minimum for $\varphi_3$ is restored and all the fields slowly roll towards their respective minimum we can see in figure \ref{fig5}. Hence, in the second stage, the field velocity is mostly dominated by $(\varphi_1,\varphi_2)$. We can that these two stages exhibit a \emph{quasi-double-inflation} like behavior where the first stage is dominated by $\varphi_3$ and the second by $(\varphi_1,\varphi_2)$.

This oscillation demonstrate a non-trivial feature to which we now turn our attention. Due to the flat metric, as expected, during the first stage, the model exhibits minor deviation from a geodesic trajectory. However, when $\varphi_3$ oscillates and the other two fields start to roll, it leads to high value of turning rate. The torsion which takes care of how much non-planar behavior is present in the trajectory, despite being small for most of the evolution, increases right after the onset of oscillation. \par

We can notice the same  even at the level of the masses of the adiabatic and entropic modes--- they switch their roles as can be seen in figure \ref{fig:pertb_mass}. The domination of entropic mass at small scales indicates the fact that at the level of perturbation, three field analysis will become unavoidable. This type of features are closely related to the production of primordial black holes at small scales -- possibly at the time of radiation domination. Besides, it can also lead to non-trivial footprint on the scalar induced gravitational waves. We plan to return to this detailed study in a companion paper. We emphasize that this trajectory is very generic for this class of models, depending on the initial condition one can easily achieve this three-field attractor. 

There exists few areas in the inflationary model and  the examples presented in table \ref{tab1}-\ref{tab2} that call for further refinement. The example in table \ref{tab3}, the vev of the blow-up modulus is calculated to be of order $\mathcal{O}(10^6)$ whereas the overall volume is $\mathcal{O}(10^5)$. However, we have made sure that the model still resides in the large volume regime throughout the inflationary evolution. Note that, this is not the case for the dS models presented in table \ref{tab1}-\ref{tab2}. The inflationary model is chosen so that the amplitude of the power spectrum is obtained without any fine-tuning or artificial scaling. Besides, the Kaluza-Klein and the string mass scales are almost comparable to each other in all the models--- pushing us to the boundary of parametric control. In all the examples de Sitter vacua presented in this paper, large value of blow-up modulus invariably implies a vanishing non-perturbative effects. Since, these effects do not play an active role in moduli stabilization, we do not face any issue. \par

The log-loop corrections are presented in a conjectured form for a K3-fibred CY, based on calculations performed in the case of a torus. In addition, since the stabilization proceeds via loop corrections, the axionic moduli remains massless because the K\"ahler potential retains the axionic shift symmetry. Subsequently, these axions can play non-trivial roles in driving several post-inflationary aspects such as reheating, and can source dark radiation and dark matter (see \cite{Cicoli:2012aq,Cicoli:2023opf} for details). Moreover, it has been emphasized in \cite{Blanco-Pillado:2009dmu} that if one considers the inflationary evolution of the axions, it usually gives rise to a greater variety of trajectories --- with cases where the axionic evolution increase the effective number of efolds. However, addressing the stabilization, evolution and dynamics of the axions are beyond the scope of this paper. Furthermore, exact computation of the values of $\chi$ and $\lambda$ is needed to unravel the robustness of the inflationary trajectories against additional string loop effects \cite{Gao:2022uop}. We save these questions for a future work.\par

%In this we have extended the previously studied moduli stabilization mechanism in fibre inflation models, while we proposed a multi-field inflation paradigm. The analysis is focused on the implications of the logarithmic loop corrections in the stabilization procedure, where the subleading $F^4$ higher derivative corrections were utilized to uplift the small flat direction. Moreover, the above method is compatible with the large volume scenario (LVS) in which where the compatification's volume is stabilized to large values of order $>\mathcal{O}(10^4)$ in accordance with the weak coupling regime.   \par

%Inflation part\par
 %An advantage of the proposed model is that it manages to have the EFT under control, despite the fact that it lacks an explicit mechanism of uplifting the vacuum to dS space. It would be worth studying the effects of the $\overline{D3}$ uplifting method \cite{Cicoli:2024bxw} in the presence of the logarithmic loop corrections. We save these questions for a future work.

\section*{Acknowledgments}

We thank George Leontaris, Sonia Paban, Pramod Shukla, and Ivonne Zavala for useful discussions and commments.

\bibliographystyle{utphys}

\bibliography{references1}

\end{document}